\documentclass[useAMS,usenatbib]{mn2e}
\usepackage[modulo]{lineno}
\usepackage{pdflscape}
\usepackage{graphicx}
\usepackage{booktabs}% http://ctan.org/pkg/booktabs
\newcommand{\tabitem}{~~\llap{\textbullet}~~}

\def\teff{T_{\rm eff}}
\def\feh{\rm[Fe/H]}
\def\logg{\log\,g}
\def\mh{\rm[M/H]}
\def\afe{\rm[\alpha/Fe]}
\def\msun{\rm M_{\odot}}
\def\dnu{\Delta\nu}
\def\num{\nu_\mathrm{max}}

\begin{document}

\title[Age dissection of the Milky Way disk(s)]{Confirming chemical clocks: asteroseismic age dissection of the Milky Way disk(s)}
\author[V.~Silva Aguirre et al.]{V.~Silva Aguirre$^1$, M.~Bojsen-Hansen$^1$, D.~Slumstrup$^1$, L.~Casagrande$^2$, D.~Kawata$^3$, 
\newauthor I.~Ciuc${\rm \breve{a}}$$^3$, R.~Handberg$^{1}$, M.~N.~Lund$^{4,1}$, J.~R. Mosumgaard$^1$, D. Huber$^{5,6,1}$, 
\newauthor J.~A.~Johnson$^{7,8}$, M.~H.~Pinsonneault$^7$, A.~M. Serenelli$^9$, D. Stello$^{10,6,1}$, J.~Tayar$^7$,
\newauthor J.~C.~Bird$^{11}$, S.~Cassisi $^{12}$, M.~Hon$^{10}$, M.~Martig$^{13}$, P.~E. Nissen$^1$, H.~W.~Rix$^{14}$, 
\newauthor R.~Sch\"onrich$^{15}$, C.~Sahlholdt$^1$, W.~H.~Trick$^{14}$, and J.~Yu$^{6,1}$\\\\
$^1$Stellar Astrophysics Centre, Department of Physics and Astronomy, Aarhus University, Ny Munkegade 120, DK-8000 Aarhus C, Denmark\\
$^2$Research School of Astronomy \& Astrophysics, Mount Stromlo Observatory, The Australian National University, ACT 2611, Australia\\
$^3$Mullard Space Science Laboratory, University College London, Holmbury St. Mary, Dorking, Surrey RH5 6NT, UK\\
$^4$School of Physics and Astronomy, University of Birmingham, Edgbaston, Birmingham, B15 2TT, UK\\
$^5$Institute for Astronomy, University of Hawaii, 2680 Wood- lawn Drive, Honolulu, HI 96822, USA\\
$^6$Sydney Institute for Astronomy (SIfA), School of Physics, University of Sydney, NSW 2006, Australia\\
$^7$Department of Astronomy, The Ohio State University, Columbus, OH 43210, USA\\
$^8$Center for Cosmology and AstroParticle Physics, 191 West Woodruff Ave, Ohio State University, Columbus, OH 43210\\
$^9$Instituto de Ciencias del Espacio (ICE-CSIC/IEEC), Campus UAB, Carrer de Can Magrans, s/n, 08193 Cerdanyola del Valles, Spain\\
$^{10}$School of Physics, University of New South Wales, NSW 2052, Australia\\
$^{11}$Department of Physics and Astronomy, Vanderbilt University, 6301 Stevenson Circle, Nashville, TN 37235, USA\\
$^{12}$INAF - Astronomical Observatory of Teramo, via M. Maggini, sn. Teramo, Italy\\
$^{13}$Astrophysics Research Institute, Liverpool John Moores University, 146 Brownlow Hill, Liverpool L3 5RF, UK\\
$^{14}$Max-Planck-Institut f\"ur Astronomie, K\"onigstuhl 17, D-69117 Heidelberg, Germany\\
$^{15}$Rudolf-Peierls Centre for Theoretical Physics, University of Oxford, 1 Keble Road, Oxford OX1 3NP, UK\\
}
\maketitle
\begin{abstract}
Investigations of the origin and evolution of the Milky Way disk have long relied on chemical and kinematic identification of its components to reconstruct our Galactic past. Difficulties in determining precise stellar ages have restricted most studies to small samples, normally confined to the solar neighbourhood. Here we break this impasse with the help of asteroseismic inference and perform a chronology of the evolution of the disk throughout the age of the Galaxy. We chemically dissect the Milky Way disk population using a sample of red giant stars spanning out to 2~kpc in the solar annulus observed by the {\it Kepler} satellite, with the added dimension of asteroseismic ages. Our results reveal a clear difference in age between the low- and high-$\alpha$ populations, which also show distinct velocity dispersions in the $V$ and $W$ components. We find no tight correlation between age and metallicity nor $\afe$ for the high-$\alpha$ disk stars. Our results indicate that this component formed over a period of more than 2~Gyr with a wide range of $\mh$ and $\afe$ independent of time. Our findings show that the kinematic properties of young $\alpha$-rich stars are consistent with the rest of the high-$\alpha$ population and different from the low-$\alpha$ stars of similar age, rendering support to their origin being old stars that went through a mass transfer or stellar merger event, making them appear younger, instead of migration of truly young stars formed close to the Galactic bar.
\end{abstract}

\section{Introduction}\label{sec:int}
%\linenumbers
Spiral galaxies such as ours contain several populations of stars comprising their bulge, disk, and halo, all with different chemical and kinematic properties capturing unique epochs of formation and the different processes that led to their specific characteristics. The disk is the defining stellar component of the Milky Way, and understanding its formation has been identified as one of the most important goals of galaxy formation theory \citep[e.g.,][]{2002ARA&A..40..487F,Rix:2013kw,BlandHawthorn:2016iq}.

The Milky Way disk has been geometrically separated into a thin and a thick component that dominate at different heights, and were identified using stellar counts more than 30 years ago \citep{1983MNRAS.202.1025G,2008ApJ...673..864J}. It is assumed that the formation history and timescale of these populations are different, and therefore stars from each geometric component of the disk should be associated with a particular chemical, kinematic, and age signature. One of the aims of Galactic archaeology is constructing the evolution history of the disk from these fossil records, which requires an accurate characterisation of the properties of stars belonging to each component.

Due to the difficulties in determining ages for faint field stars based purely on spectroscopic or photometric information, most studies of the Milky Way disk have focused on identifying different populations in the solar neighbourhood using chemistry and kinematics \citep[e.g.,][just to name a few]{1998A&A...338..161F,Soubiran:2003ev,Bensby:2003fk,Navarro:2011iw,Ramirez:2013cb,RojasArriagada:2016eq,AllendePrieto:2016hn}. It is expected that the geometric thin and thick disks can be separated using space velocities as the thick component should be kinematically hotter than the thin disk population, and given their different timescales of formation they should present different trends in $\alpha$-abundance patterns that would make them also identifiable in chemical space. However, neither the chemical nor the kinematic criterion seem to clearly separate the disk components because of the substantial overlap between the phase-space distribution of stars belonging to the high- and low-$\alpha$ abundance sequences \citep[e.g.,][]{Feltzing:2008fd,Schonrich:2009ee,Adibekyan:2011ec,Bensby:2014gi}. Indeed, the radial and vertical structure of chemically selected disk components varies intrinsically with abundance, especially among low-$\alpha$ stars \citep{Bovy:2012du,2016ApJ...823...30B}. High-resolution spectroscopic surveys suggest a bimodal distribution of $\alpha$-abundances in disk stars \citep[e.g.,][]{Anders:2014jj,2015ApJ...808..132H}, although the very existence of a chemically distinct disk has been challenged based on chemo-dynamical models and low-resolution spectroscopic observations \citep{Schonrich:2009ee,Loebman:2011bf}.

The reason for the mixing and contamination between chemical and kinematic samples is probably related to stars in the solar neighbourhood being born at different galactocentric radii and migrating to their current position by dynamical processes such as radial migration \citep[e.g.,][]{2002MNRAS.336..785S,2008ApJ...684L..79R,Schonrich:2009ci,Minchev:2013ko,Grand:2015kf}. Viable formation scenarios for the Milky Way disk include inside-out and upside-down formation \citep[e.g.,][]{1989MNRAS.239..885M,2006ApJ...639..126B,Bird:2013iw,2016ApJ...823..114N,Schonrich:2017gj}, although there are suggestions that the old component of the disk formed outside-in instead \citep{Robin:2014ky}. A natural step forward to gain further insight about these processes would be to also identify stellar populations within the thin and thick disk based on the chronology of formation events, effectively dissecting the disk by adding the age dimension.

Using a very local ($\sim25$~pc) but volume complete sample, \citet{1998A&A...338..161F,Fuhrmann:2011em} suggested that a chemical dissection of disk would also result in a clean age separation of the components: the high-$\alpha$ sequence should be occupied by stars older than about 10~Gyr while the low-$\alpha$ population is expected to be composed by stars younger than about 8~Gyr. A transition between the $\alpha$-abundances sequences should occur at 10~Gyr, and the majority of the thick disk stars are expected to have an age of about 12~Gyr. This age gap still needs to be confirmed, and the advent of asteroseismology as a tool for Galactic archaeology promises to test this paradigm by determining precise masses and ages for thousands of stars in distant regions of the Milky Way \citep[e.g.,][]{Miglio:2013hh,Casagrande:2014bd,2016MNRAS.455..987C,2017A&A...597A..30A,Rodrigues:2017fj}.

Among the efforts combining ground-based follow-up observations and oscillations data, the APOKASC catalogue \citep[][]{Pinsonneault:2015kd} comprises a sample of almost 2000 red giant stars observed by APOGEE \citep{Majewski:2017ip} that have asteroseismic detections in their frequency power spectrum obtained with the {\it Kepler} satellite. \citet{Martig:2016ck} showed that the carbon and nitrogen abundances in these giants, altered in their photospheres during the first dredge-up, could be calibrated using the masses derived from asteroseismology to provide a spectroscopic measurement of (implied) ages with a precision of $\sim0.2$~dex and an accuracy below 0.1~dex. Coupled to data driven approaches such as {\it The Cannon} \citep{2015ApJ...808...16N} they allow us to map the age distribution of thousands of red giants spanning up to $\sim5$~kpc \citep{2016ApJ...823..114N,Ho:2017ed} and have showed that the high-$\alpha$ stars are predominately old (and much older than low-$\alpha$ stars) throughout the disk.

While methods like {\it The Cannon} are capable of extending the results of asteroseismology to stars that have no oscillations data, the precision in their derived ages is significantly lower ($\sim0.2$~dex) than that obtained by asteroseismic analysis of red giants \citep[e.g.,][]{2016MNRAS.455..987C,2017A&A...597A..30A}. In this paper we re-examine the APOKASC catalogue using the most up-to-date asteroseismic observations and introduce the age dimension in chemical studies of the Galactic disk. The combination of asteroseismic inference and spectroscopic information allows us to determine precise stellar properties as well as kinematic and chemical patterns of stars, and to chemically dissect the Milky Way disk including age information in a sample spanning distances up to 3~kpc from the Sun.
\section{The APOKASC sample}\label{sec:stars}
\subsection{Photometric, spectroscopic, and asteroseismic data}\label{ssec:data}
\begin{figure}
\begin{center}
\includegraphics[width=\linewidth]{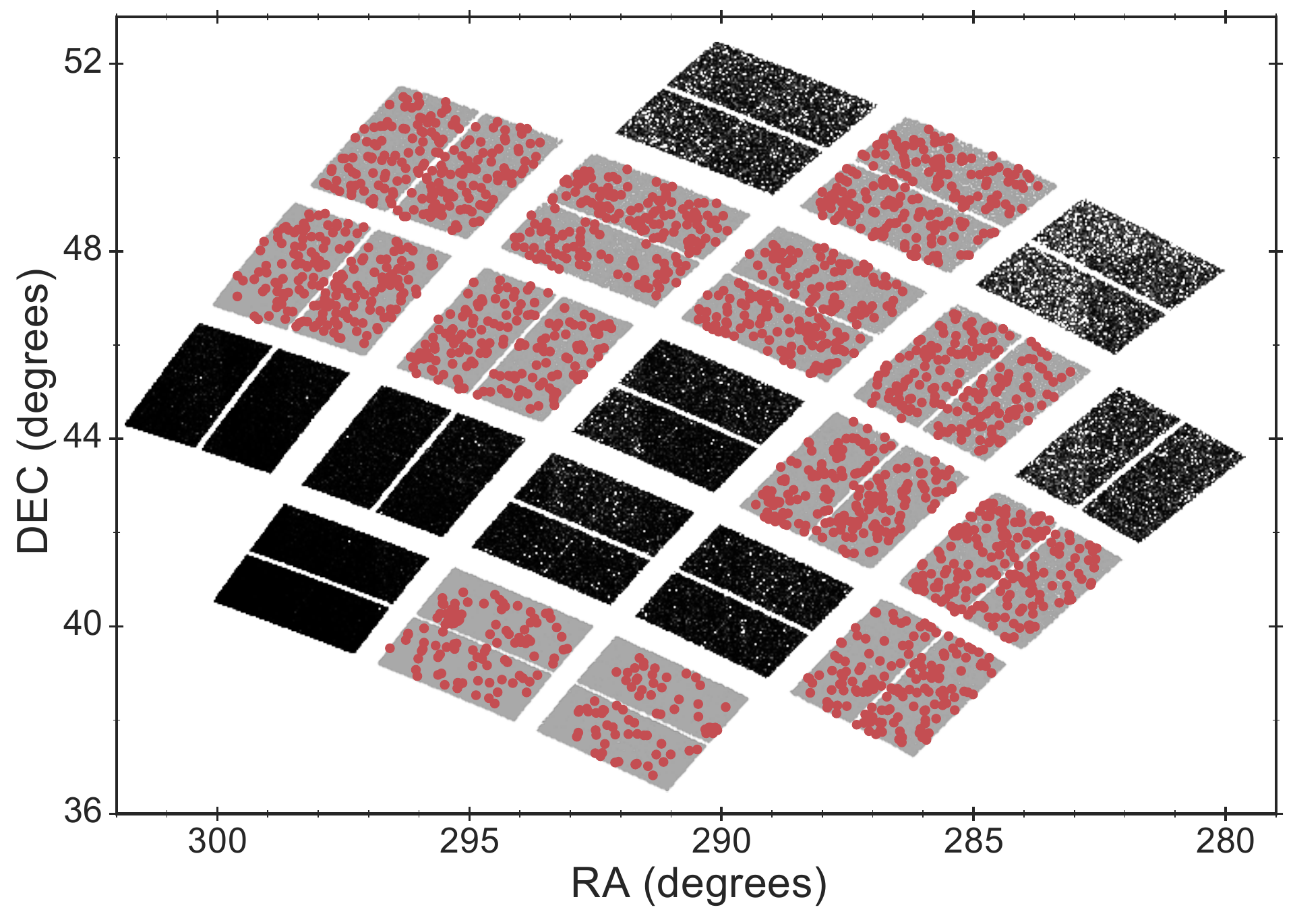}
\caption{Position in the sky of the {\it Kepler} field of view. Stars with asteroseismic detections in the APOKASC14 sample are shown in red, while grey (black) symbols depict all stars from the 2MASS catalogue falling in the same (other) CCD of the {\it Kepler} spacecraft.}
\label{fig:Kep_field}
\end{center}
\end{figure}
The set of stars considered in this study are the 1989 red giants comprising the first combined APOGEE and {\it Kepler} catalogue, in which the {\it Kepler} spacecraft detected oscillations during its nominal mission. The position of the sample in the {\it Kepler} field of view is shown in Fig.~\ref{fig:Kep_field}. From the {\it Kepler} Input Catalogue \citep[KIC][]{Brown:2011dr,Huber:2014dh} we retrieved observations for all our targets in the $griz$ filters, which we converted to standard Sloan magnitudes using the transformation derived by \citet{Pinsonneault:2012he}. Infrared observations in the $JHK_S$ bands were extracted from the 2MASS Point Source Catalogue \citep{2006AJ....131.1163S}.

The original APOKASC compilation presented by \citet{Pinsonneault:2015kd} contained asteroseismic information based on observations from the nominal {\it Kepler} mission spanning quarters Q0 to Q8 (two years of data). In total, {\it Kepler} acquired 17 quarters of data after four years of observations which are now available for analysis. Thus, we used the updated global seismic parameters $\langle\dnu\rangle$ (average large frequency separation) and $\num$ (frequency of maximum power) from Yu et al. (2017, in preparation). Among the many breakthroughs of space-based asteroseismology is the possibility of discriminating between stars in their first ascent up the red giant branch and those that have already ignited helium in their core based solely on their observed pulsation spectra \citep{Bedding:2011il,Mosser:2012jy}. We adopt the revised evolutionary classifications from \citet{Hon:2017dz}\footnote{Here we adopt updated results from the technique by \citet{Hon:2017dz} based on a new and improved neural network model (Hon et al. in preparation)}, \citet{Vrard:2016jr}, and \citet{Stello:2013jz}.

We use spectroscopic parameters from the 13th data release (DR13) of the Sloan Sky Digital Survey \citep[SDSS,][]{2016arXiv160802013S}, including metallicity, $\alpha$-abundances, effective temperatures, and heliocentric radial velocities. The uncertainties in metallicity reported in this compilation correspond to the internal precision and are of the order of $\sim$0.03~dex. In an effort to better estimate systematic uncertainties, we added in quadrature the median difference between APOGEE results in clusters and the standard literature values as reported in Table~3 of \citet{Tayar:2017du} (corresponding to 0.09~dex). Atmospheric properties were available for 1984 of the targets, and we further neglected 5 targets with metallicities consistent with $\feh<-1$, as the reliability of asteroseismic determination of stellar masses in the metal-poor regime could be slightly biased towards too high values \citep[see e.g.,][for a discussion]{Epstein:2014ct,Miglio:2016ba}.
\subsection{Determination of stellar properties}\label{ssec:basta}
We determined stellar properties for the APOKASC sample combining the photometric, spectroscopic, and asteroseismic observables using the BAyesian STellar Algorithm \citep[{\tt BASTA},][]{2015MNRAS.452.2127S,SilvaAguirre:2017eh}. The procedure for deriving all quantities is divided into two steps, which we describe in the following.

In the first step we determine all physical properties of the stars in the sample such as mass, radius, luminosity, and age, using the so-called {\it grid-based} method. In its essence, this approach compares all observed quantities with predictions from theoretical models of stellar evolution. Thus, the only requirement to apply the {\it grid-based} method is a set of tracks or isochrones containing all observed quantities and covering the necessary parameter space. We use a set of BaSTI isochrones \citep{Pietrinferni:2004im} including the effects of overshoot in the main sequence and semiconvection in the core He-burning phase that have been extended in metallicity coverage especially for asteroseismic studies \citep[see][for a description]{SilvaAguirre:2013in}.

The two atmospheric properties we fit are the effective temperature $\teff$ and bulk metallicity $\mh$. The latter is determined from the DR13 measurements of $\feh$ and $\afe$ following the prescription of \citet{Salaris:1993iv}. We complement this information with the global asteroseismic parameters $\langle\dnu\rangle$ and $\num$, that must first be determined in our set of isochrones. To do this we consider the asteroseismic scaling relations \citep{Ulrich:1986ge,Brown:1991cv}, which can be written as:
\begin{equation}\label{eqn:sca_dnu} 
\left(\frac{\langle\dnu\rangle}{\langle\dnu_\odot\rangle}\right)^{2} \simeq \frac{\bar{\rho}}{\bar{\rho}_\odot}\,, 
\end{equation}
\begin{equation}\label{eqn:sca_num} 
\frac{\num}{\nu_{\mathrm{max},\odot}} \simeq \frac{M}{\msun}{\left(\frac{R}{{\rm R_\odot}}\right)^{-2}} \left(\frac{\teff}{T_{\mathrm{eff},\odot}}\right)^{-1/2}\,. 
\end{equation}
Here $\langle\dnu_\odot\rangle$ and $\nu_{\mathrm{max},\odot}$ are the solar reference values and they depend on the asteroseismic pipeline used to extract the pulsation information from the {\it Kepler} light-curves. For our adopted set of asteroseismic observables they correspond to $\langle\dnu_\odot\rangle=135.1~\mu$Hz and $\nu_{\mathrm{max},\odot}=3090~\mu$Hz \citep{Huber:2011be}.

Equations~\ref{eqn:sca_dnu} and~\ref{eqn:sca_num} can be used to estimate the theoretical values of $\langle\dnu\rangle$ and $\num$ for any point along an evolutionary track or isochrone. Testing the accuracy of the asteroseismic scaling relations is currently a very active field of research, and there is strong evidence of a metallicity, effective temperature, and evolutionary phase-dependent offset in the large frequency separation relation \citep[cf. Eq.~\ref{eqn:sca_dnu}, see][]{White:2011fw, 2016ApJ...822...15S,Guggenberger:2016if}. In {\tt BASTA} we apply the correction by Serenelli (2017, in preparation) to this equation, which has been shown to reproduce a number of classical age determinations (e.g., turn-off fitting, eclipsing binaries, white dwarf cooling curve) in the open clusters M67 \citep{Stello:2016hu} and NGC~6819 \citep{2016MNRAS.455..987C}. Several tests carried out on the scaling relations suggest they are accurate to a level of a few percent \citep[see e.g.,][]{Huber:2012iv,SilvaAguirre:2012du,White:2013bu,Miglio:2013fb,Miglio:2016ba,2016ApJ...832..121G,Huber:2017fg}.

When the evolutionary phase of each target is known from the asteroseismic analysis of the power spectrum we impose it in the analysis as a Bayesian prior, otherwise all possible stages of evolution are taken into account in {\tt BASTA} when constructing the probability density function for determining the stellar properties. This results in unclassified targets normally having larger statistical uncertainties than RGB or clump stars (see section~\ref{ssec:age_unc} below), but they represent a small fraction of our sample and therefore do not significantly affect the conclusions of our study.

The current version of {\tt BASTA} includes a feature to compute distances based on bolometric corrections from observed magnitudes in different bandpasses and an extinction map, following the procedure outlined by \citet{Rodrigues:2014it}. Thus, the second step in our {\it grid-based} method consists of using the surface gravity determined by {\tt BASTA} and combining it with the reddening value from the KIC catalogue, the $grizJHK_S$ magnitudes, and the spectroscopic $\teff$ and $\feh$ to compute the bolometric correction using the synthetic photometry of \citet{Casagrande:2014ez}. We then determine the distance modulus in each of the seven bandpasses and use the median value to extract the bolometric magnitude as described by \citet{Torres:2010gd}, and obtain the distance to the star. From the derived distance and coordinates of the target, we recompute the extinction value using the \citet{Green:2015cf} reddening map. This updated interstellar extinction is combined again with the surface gravity, photometric and spectroscopic measurements to obtain a new bolometric correction, and therefore a new distance estimate. The process is repeated until extinction changes by less than 0.001 mag between iterations and thus convergence is reached (normally only two iterations are needed). Our results are in excellent agreement with those from other asteroseismic pipelines used to determine distances such as the Bellaterra Stellar Properties Pipeline \citep[][]{Serenelli:2013fz} and the PARAM code \citep{DaSilva:2006be,Rodrigues:2014it}. A comparison between the estimates from these pipelines can be found in Fig.~2 of \citet{Huber:2017fg}.
\subsection{Sample completeness}\label{ssec:selfun}
We describe in the following the pruning procedure applied to this sample according to data availability and quality. For the 1979 stars with stellar properties determined, we collected proper motions from the first data release of Gaia \citep{Lindegren:2016gr,GaiaCollaboration:2016gd} or the UCAC-4 catalogue \citep{Zacharias:2013cf} according to availability. The latter were further pruned following the quality check procedure described by \citet{Anders:2014jj} (see their section 3.2), yielding a total of 1838 stars with reliable astrometric solutions. Using our asteroseismically determined distances we followed \citet{1987AJ.....93..864J,Bensby:2003fk} to determine the (right-handed) Galactic space-velocity components $UVW$ in the Local Standard of Rest as defined by \citet{Schonrich:2010kt}. We performed Gaussian fits to determine the standard deviation in each velocity component of the full sample, and as a further data-quality control we selected only the targets with uncertainties in their individual velocities below half of the computed standard deviation, resulting in a sample of 1593 stars with precise kinematic information available. Finally, we removed 3 stars for which our derived stellar properties resulted in fractional age uncertainties larger than 1.

Our sample of 1590 stars selected in this manner is a subset of the more than 15,000 oscillating red giants detected by {\it Kepler}, and the fact that this set and not a different one is available for our study is the result of a combination of criteria largely based on plate availability for observations with the APOGEE telescope. Thus, there is no reason to believe {\it a priori} that the stars comprising the APOKASC sample (or our pruned sub-selection of it) are representative of the physical and kinematic characteristics of the true underlying population of red giants in that direction of the sky. Any inference directly drawn from this sample regarding e.g. its age or composition distribution carries the risk of being biased unless we properly correct for the selection function. To gauge this effect, we follow \citet{2016MNRAS.455..987C} and take a two-step approach where we first correct for the selection of oscillating giants with available APOGEE spectra, and after that for the target selection effects of the {\it Kepler} spacecraft as a function of distance (see section~\ref{sec:tarprob} below).
\begin{figure}
\begin{center}
\includegraphics[width=\linewidth]{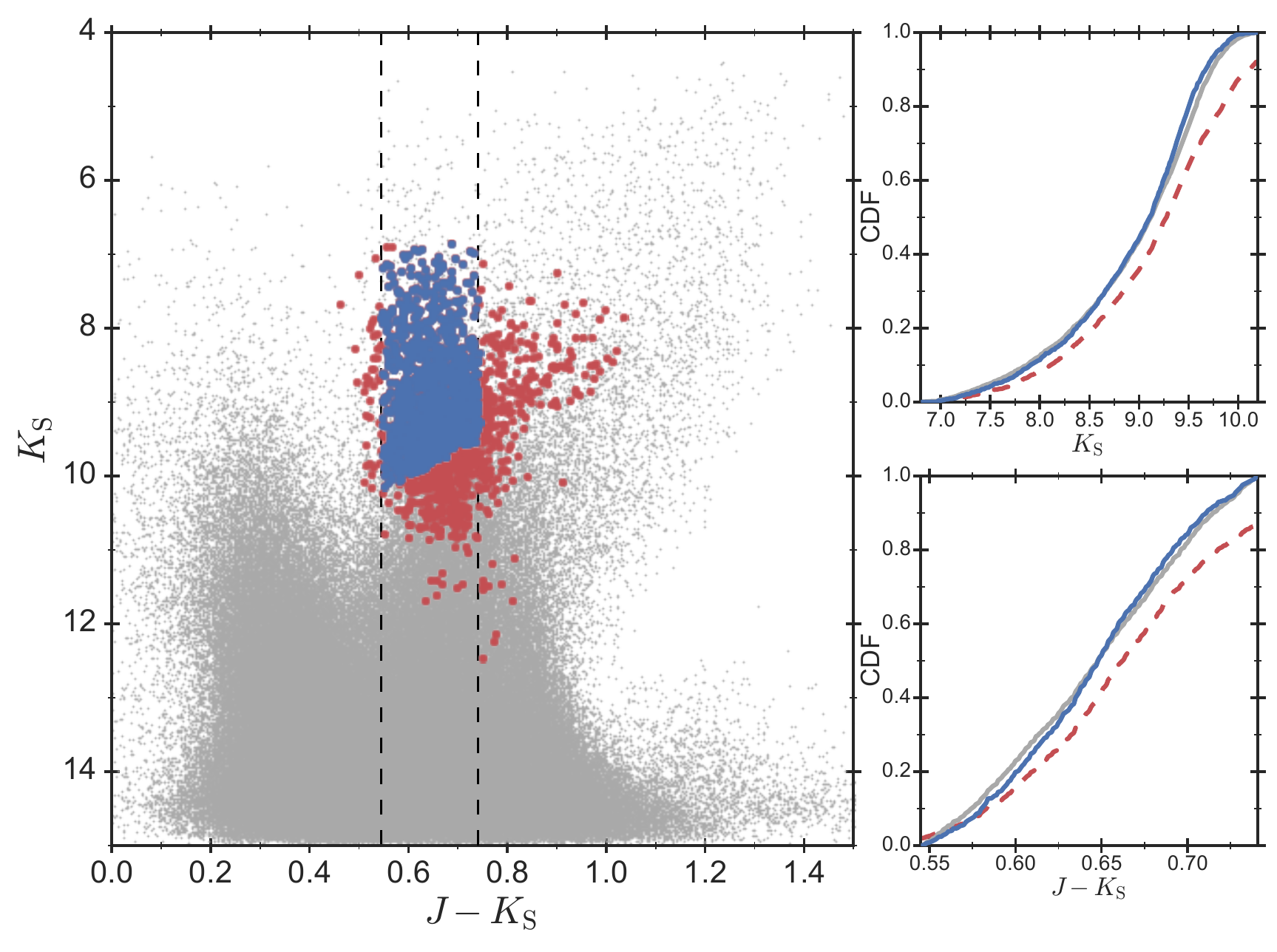}
\caption{Selection function for a photometrically complete sample. {\it Left}: colour-magnitude diagram of all stars from 2MASS (grey) falling within the same CCD's as the APOKASC sample (shown in red). The stars fulfilling the photometric completeness criteria are shown in blue. {\it Right}: cumulative distributions in magnitude and colour of the 2MASS sample (grey solid), the APOKASC stars (red dashed), and the pruned complete sample (blue solid). See text for details.}
\label{fig:2mass_sel}
\end{center}
\end{figure}

To assess the potential biases of the selected sample of 1590 targets, we need to compare their intrinsic properties to an unbiased set of stars in the same field of view. Fortunately, the {\it Kepler} field has available 2MASS photometry that we can use to quantify these effects. The main panel in Fig.~\ref{fig:2mass_sel} shows a colour-magnitude diagram of all stars in the 2MASS catalogue falling within the same CCDs of {\it Kepler}'s field of view as our targets (c.f Fig.~\ref{fig:Kep_field}). We repeated the procedure outlined in \citet{2016MNRAS.455..987C} to find the magnitude and colour ranges where our sample of stars and the underlying 2MASS population are indistinguishable from each other. The resulting cumulative distributions are shown in the right panels of Fig.~\ref{fig:2mass_sel}, where we have removed stars and applied the K-Sample Anderson-Darling test until the null hypothesis that both samples are drawn from the same parent distribution cannot be rejected with a significance level greater than 1\% for both $J-K_\mathrm{S}$ and $K_\mathrm{S}$. This statistical test is similar to the commonly applied Kolmogorov-Smirnoff test but much more sensitive at the edges of the distribution. The procedure leaves us with a final set of 1197 stars that are fully representative in colour and magnitude of the underlying stellar population in the {\it Kepler} line of sight (see Table~\ref{tab:obs}). The updated evolutionary classifications for this sample used in this paper yield a total of 422 stars in the RGB and 751 in the clump phase, while it is not possible to assign an evolutionary status to the remaining 24 giants.
\begin{table}
\centering
\caption{Number of stars considered in this study according to data availability and quality control. See text for details.}
\label{tab:obs}
\begin{tabular}{lc}
\hline
\multicolumn{2}{|c|}{APOKASC Catalogue} \\
\hline
\smallskip
Total number of stars & 1989 \\
\hline
\smallskip
{\tt BASTA} stellar properties & 1979 \\
\smallskip
Available proper motions & 1838 \\
\smallskip
Galactic velocities cut & 1593 \\
\smallskip
Seismic uncertainties cut & 1590 \\
\hline
\smallskip
Colour-magnitude complete sample &  1197 \\
\smallskip
\tabitem RGB &  422 \\
\smallskip
\tabitem Clump & 751 \\
\smallskip
\tabitem Unclassified & 24 \\
\hline
\end{tabular}
\end{table}
\subsection{Final uncertainties}\label{ssec:age_unc}
The procedure outlined in section~\ref{ssec:basta} results in a set of stellar properties and their associated errors that takes into account the observational uncertainties but does not yet consider systematics based on the input physics of our stellar models. With this in mind, we derived a second set of properties for our sample with {\tt BASTA} using the same input parameters but a grid of stellar models including the effects of mass-loss. The latter was implemented following the Reimers prescription \citep{Reimers:1977ts} with an efficiency of $\eta=0.4$, representing the extreme of the commonly adopted values around $\eta=0.1-0.2$ supported by various observations \citep[e.g.,][]{Miglio:2012dm,Origlia:2014ew,Miglio:2016ba}. This allows us to determine a conservative estimate of the age uncertainty introduced to our results by mass-loss.

To determine the final uncertainties in the stellar properties we first consider results from our standard grid of models that includes overshooting, semiconvection, and no mass-loss. In {\tt BASTA} we obtain the posterior probability density function and we add in quadrature the 16 and 84 percentiles to determine the statistical uncertainty in our derived stellar properties. We add in quadrature to these uncertainties a systematic component calculated as half the difference between the standard values and those obtained with the grid with very efficient mass-loss. After including this effect our resulting stellar properties have median uncertainties of the order of 2.7\% (radius), 3.3\% (distance), 7.0\% (mass), and 28.5\% (age), in line with those obtained by e.g., \citet{Casagrande:2014bd,2016MNRAS.455..987C,2017A&A...597A..30A,Rodrigues:2017fj} using similar approaches.
\begin{figure}
\begin{center}
\includegraphics[width=\linewidth]{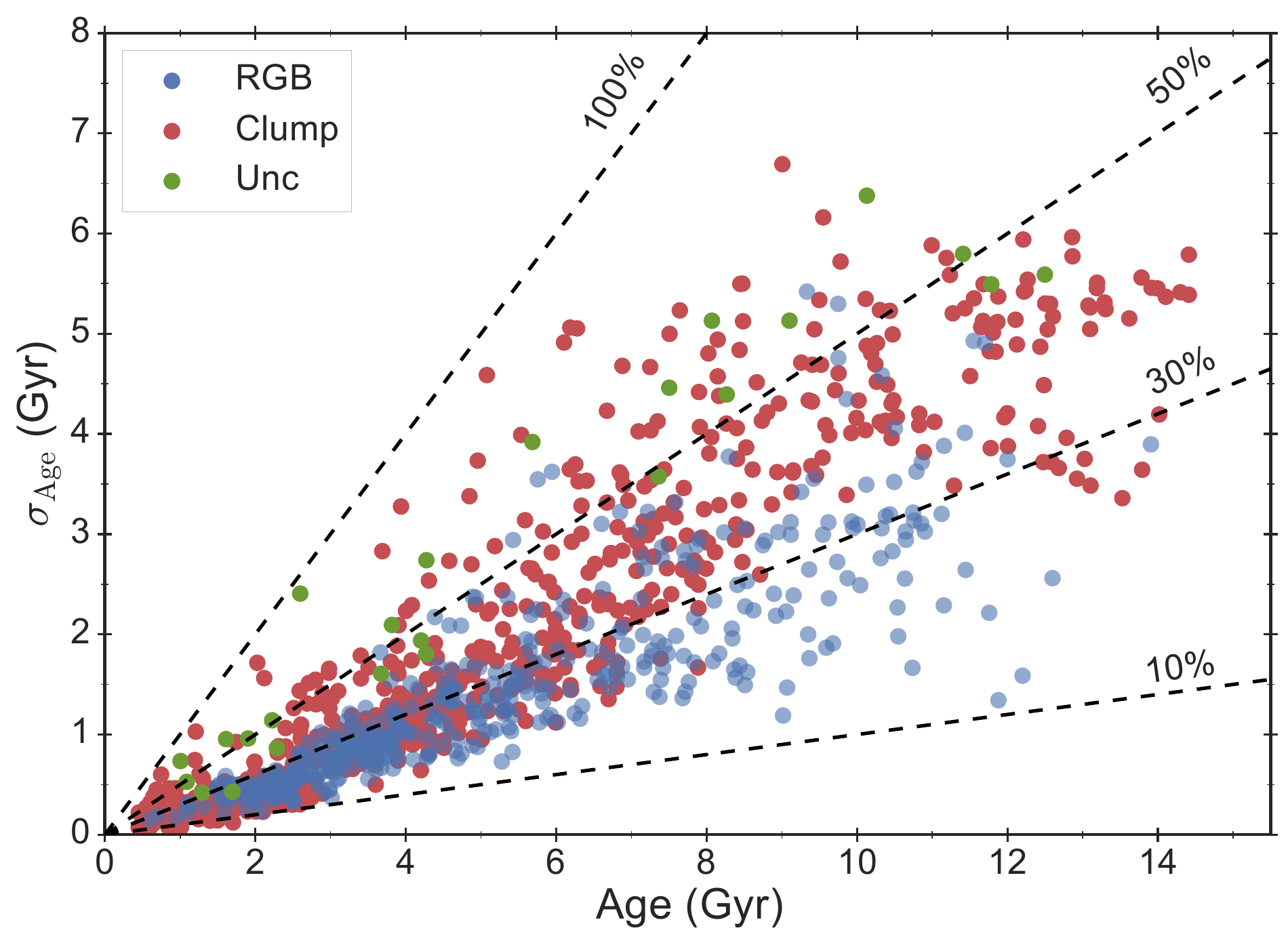}
\caption{Final age uncertainties obtained for the colour-magnitude complete sample, colour-coded according to their asteroseismic evolutionary classification. Dashed lines mark the 10\%, 30\%, 50\%, and 100\% fractional uncertainty levels.}
\label{fig:ages_unc}
\end{center}
\end{figure}

The age uncertainties for our sample are plotted in Fig.~\ref{fig:ages_unc} where it can be seen that, as age increases, clump stars have larger fractional uncertainties than RGB targets of the same age. This is the result of including mass-loss in our calculations, which in the Reimers prescription is most efficient for low-mass stars and towards the tip of the RGB. The unclassified stars in the sample have in general larger uncertainties than their clump and RGB age counterparts as a consequence of the flat Bayesian prior used for their evolutionary state (as mentioned in section~\ref{ssec:basta}).
\begin{figure}
\begin{center}
\includegraphics[width=\linewidth]{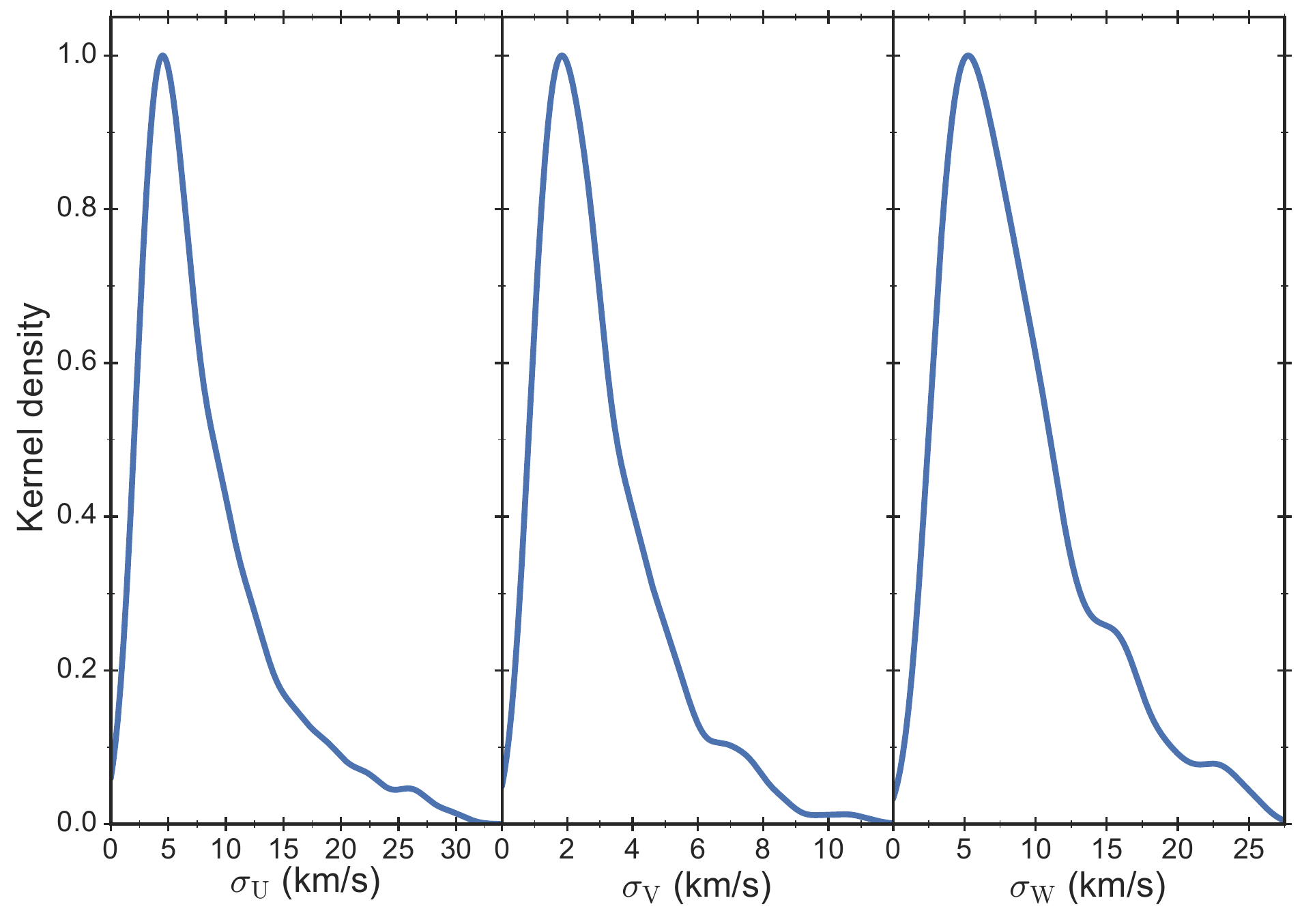}
\caption{Uncertainties in the Galactic velocities derived for the sample. The kernel density estimates have been computed using a Gaussian kernel and scott's rule for determining the bandwidth.}
\label{fig:Gvel_unc}
\end{center}
\end{figure}

Figure~\ref{fig:Gvel_unc} depicts the resulting uncertainties in the Galactic velocity components for our complete sample of stars. The  kernel density estimates (KDE) show that the combination of small uncertainties in our derived distances and the applied kinematic pruning criteria (c.f section~\ref{ssec:selfun}) results in {\bf precise} space velocities. We note that the uncertainties in the $V$-velocity component are smaller than those in $U$ and $W$ because the {\it Kepler} field lies approximately in the direction of the $V$-velocity component; therefore its uncertainty is mainly determined by the precision in the radial velocity measurements from spectroscopy. On the other hand, the absolute errors in $U$ and $W$ are mostly dominated by the uncertainties in proper motions.
\section{Correcting for target selection effects}\label{sec:tarprob}
Our colour and magnitude complete sample of giants comprises metallicities between $-1.0<\feh<0.5$, ages from $\sim0.4$ to $\sim14$~Gyr, gravities within $3.3<\logg<2.0$, and distances from $\sim300$ to $\sim3000$~pc. There are many thousands of red giant stars in the {\it Kepler} field of view with stellar properties falling within these ranges but only 1197 of the ones in the subsample studied here have detections of oscillations reported. As a result, our oscillating targets represent only a fraction of the stars that could have pulsations detected by {\it Kepler} at a given combination of age, metallicity, and distance, and therefore it is possible that each star in our photometrically complete sample is either over-or-underrepresented.

To take into account the stellar population biases due to our particular target selection applied in section~\ref{ssec:selfun}, we follow the approach presented by \citet{2016MNRAS.455..987C}, who corrected for the {\it Kepler} target selection bias generating a multidimensional synthetic data cube. The underlying idea is that at a given distance, there is a ratio between the number of stars where oscillations are detected and the total number of stars where oscillations could have been detected if all stars would have been selected by the {\it Kepler} science team for observations. This ratio varies as a function of distance because the observed magnitude of stars changes, making them fall in or out of our derived colour-magnitude completeness region defined in section~ \ref{ssec:selfun}. Thus, a high value of the ratio represents a high fraction of stars being selected for observations with {\it Kepler} were oscillations can be detected.

We created a multi-dimensional data cube in age and metallicity using the set of BaSTI isochrones with overshoot and no mass-loss, and randomly populated it by generating synthetic stellar samples from a Salpeter initial mass function. We assigned apparent magnitudes to each synthetic star by running over the distance dimension. Once the probabilities for each synthetic target are calculated as the ratio between the number of stars satisfying our completeness criteria and the total number of stars at a given distance, the probabilities for our actual targets are determined using interpolation along the 3-axis of the data cube. The most underrepresented stars have probabilities consistent with zero, which we then determine by taking the median of 10,000 Monte Carlo realisations of the interpolation assuming a gaussian distribution of their uncertainties in age, metallicity, and distance.
\begin{figure}
\begin{center}
\includegraphics[width=\linewidth]{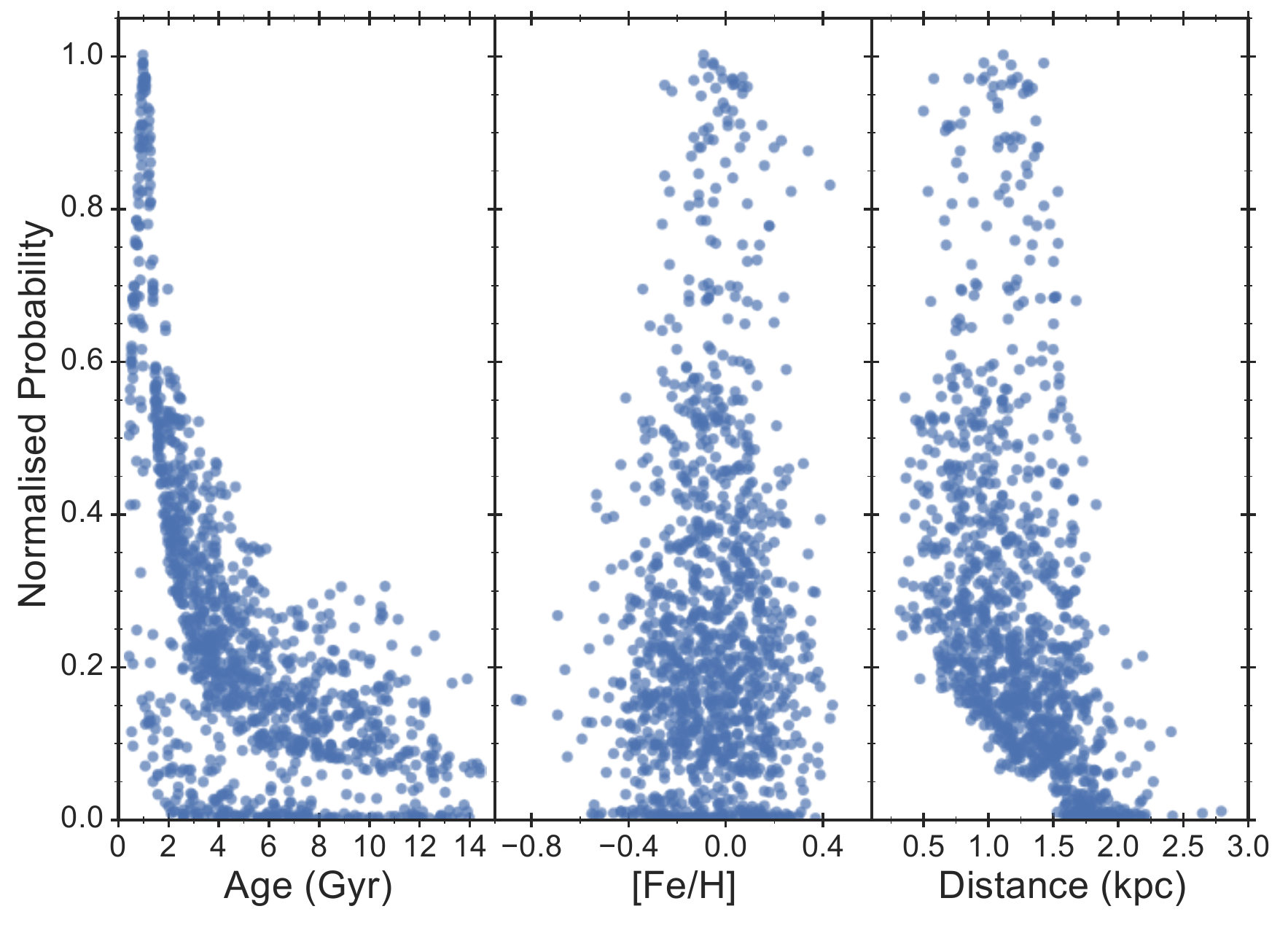}
\caption{Normalised probability of stars being observed by the {\it Kepler} satellite at different ages, metallicities, and distances.}
\label{fig:targ_sel}
\end{center}
\end{figure}

Figure~ \ref{fig:targ_sel} shows the probabilities resulting from our target selection procedure as a function of age, metallicity, and distance. Stars with lower probability are disfavoured by the {\it Kepler} target selection, while those with high probability values are the most likely to have oscillations detected. The resulting trends can be understood in terms of the stellar properties governing the detectability of pulsations, in particular the oscillation amplitudes that scale proportionally to the stellar luminosity \citep{Kjeldsen:1995tr}. For example, it is clear from Fig.~ \ref{fig:targ_sel} that young stars are favoured over old stars. The reason is that at a given metallicity and distance more massive (younger) stars are intrinsically brighter than their less massive (older) counterparts, resulting in an overrepresentation of young stars in our sample. Similarly, stars with the lowest probabilities are scattered across the age and metallicity range but restricted to distanced beyond $\sim1.5$~kpc, as detecting the intrinsic variations in brightness characterising the oscillations is harder for the furthest away targets.

The procedure outlined above allows us to gauge the effect of the {\it Kepler} target selection in our results, and we correct for this effect in all the following results by weighting each star in our fits and distributions by the inverse of the obtained probability and its corresponding uncertainty. In practise this means that we resample our stars a number of times proportional to the logarithm of their inverse individual probability. Before closing this section, we note that different approaches for determining this correction such as population synthesis and galaxy modelling were explored by \citet{2016MNRAS.455..987C}, who found consistent results across all methods. Similarly we have tested the impact of changes in the initial mass function by adopting a uniform distribution in masses instead of a Salpeter IMF when populating the isochrones, and found no significant difference in the distribution of probabilities (see appendix~\ref{app_tsel}). Our choice of the data cube for correcting target selection effects is guided by having the least model-dependent approach, but we emphasise that adopting a different methodology does not have a significant impact in our derived results.
\section{Separating disk components}\label{sec:separ}
After the data pruning, sample completeness, and target selection effects procedures described in sections~\ref{sec:stars} and~\ref{sec:tarprob}, we have a set of 1197 red giants with precise physical, chemical, and kinematic properties which is representative of the population of giants in the {\it Kepler} line-of-sight. We now turn our attention into the analysis of the properties of this sample and its implications for the formation and evolution of the Galactic disk.
\begin{figure}
\begin{center}
\includegraphics[width=\linewidth]{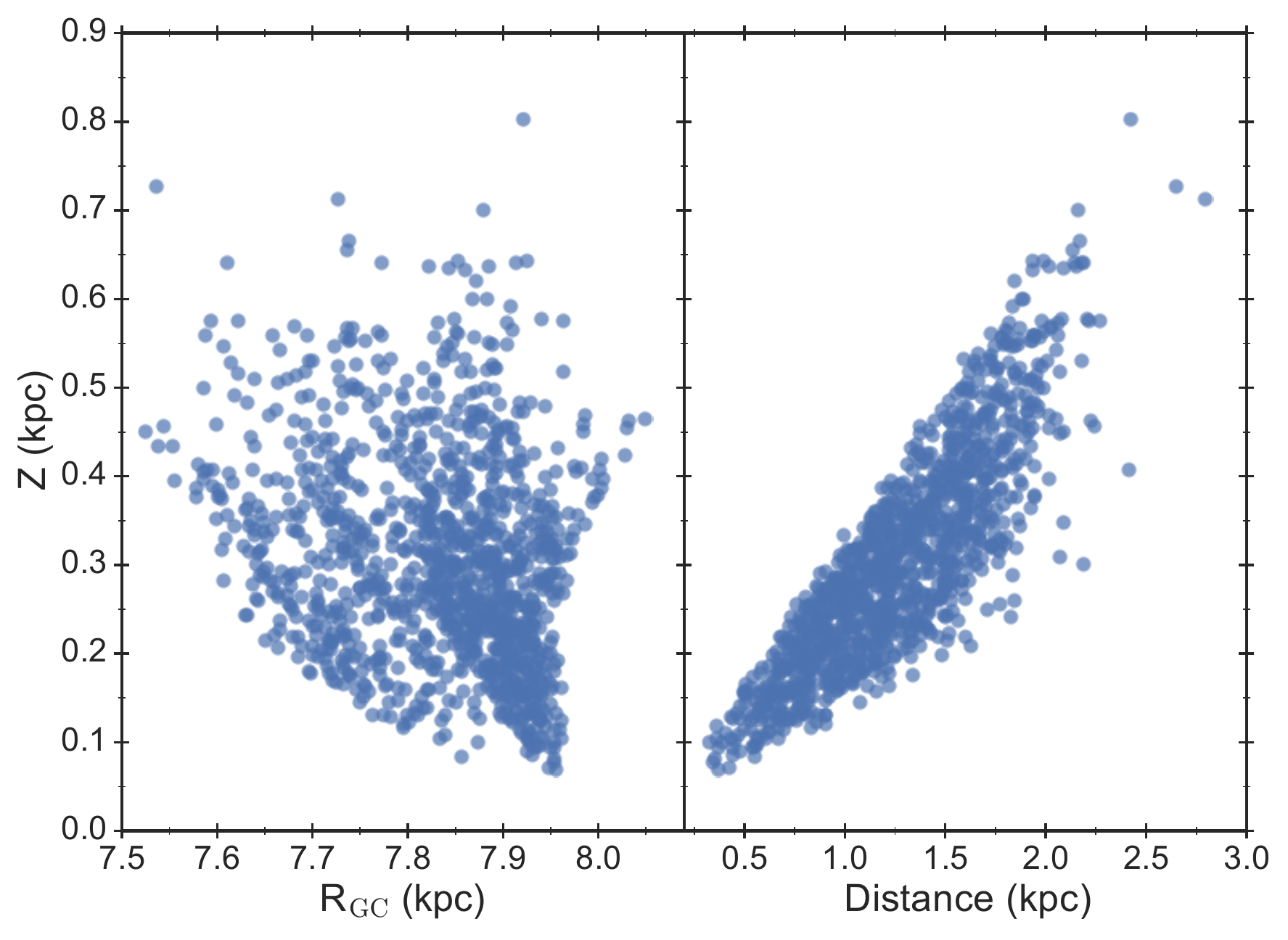}
\caption{Position in the Galaxy of the complete sample determined assuming a solar position of R$_{\mathrm{GC,}\odot}=8$~kpc. {\it Left:} Galactocentric radii as a function of height above the plane. {\it Right:} distance from the Sun. See text for details. }
\label{fig:gal_pos}
\end{center}
\end{figure}

The positions in the Galaxy of the stars in our complete sample are shown in Fig.~ \ref{fig:gal_pos}. Due to the location of the {\it Kepler} field at Galactic longitude of $l\simeq74^{\circ}$ and the pencil-beam shape nature of the survey, all targets are roughly located at the same galactocentric radii as the Sun R$_{\mathrm{GC,}\odot}\simeq8$~kpc, giving us a fully representative population of red giants in the solar annulus. Our targets also span distances up to $\sim2$~kpc and almost 1~kpc above the plane, probing much further than commonly studied samples in e.g.,. the solar neighbourhood which are volume-complete to 25~pc \citep[][]{Fuhrmann:2011em} or $\sim40$~pc \citep{2004A&A...418..989N,Casagrande:2011ji}.
\begin{figure}
\begin{center}
\includegraphics[width=\linewidth]{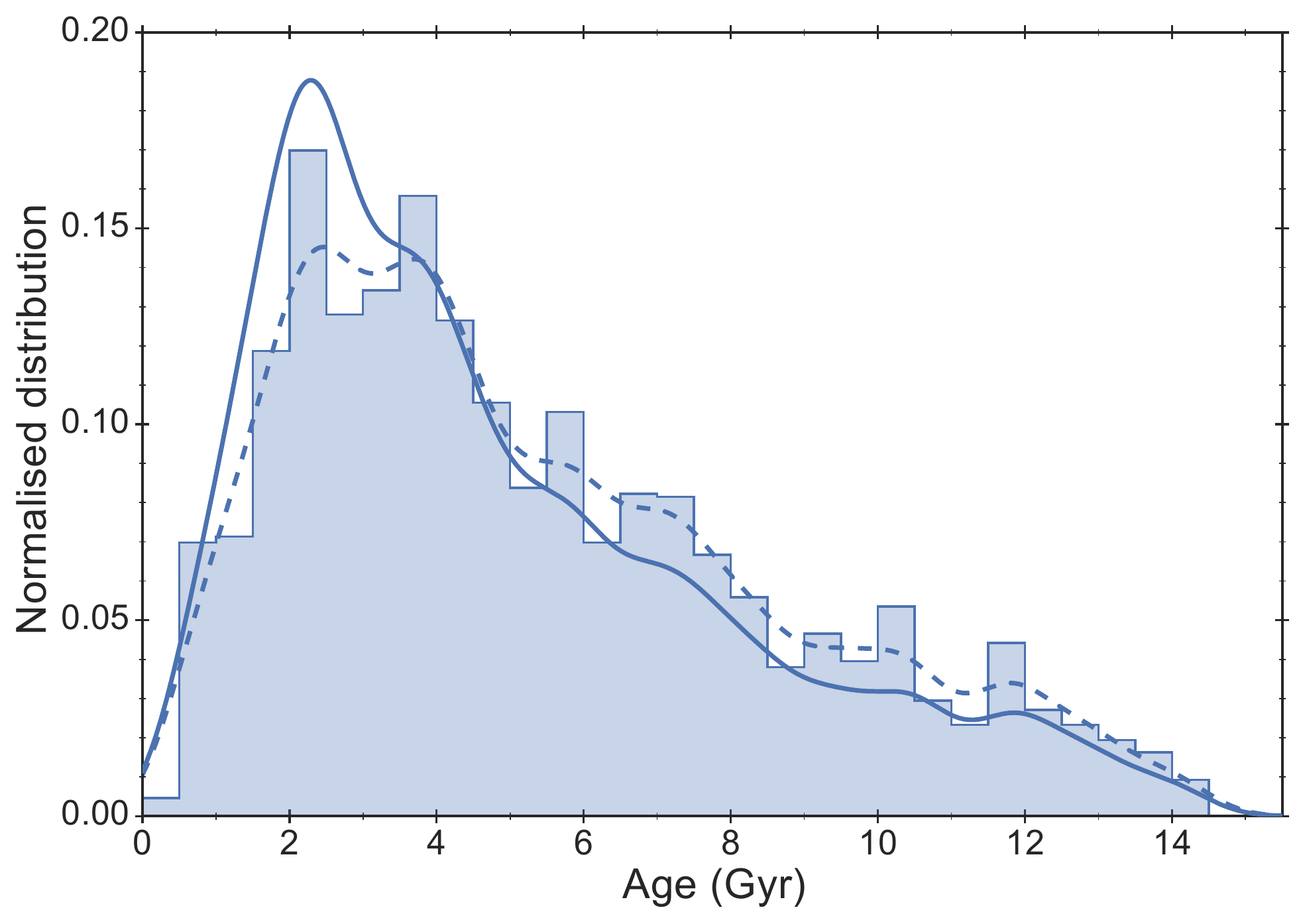}
\caption{Age histogram of the colour-magnitude complete sample (shaded region). Dashed curve depicts the unweighted kernel density estimate of the distribution calculated with a Gaussian kernel, while the solid curve takes into account the age uncertainties as weights. All distributions are normalised such as the integral over the range is unity. See text for details.}
\label{fig:Agedist}
\end{center}
\end{figure}

Figure~ \ref{fig:Agedist} shows the age histogram of the sample constructed with equally spaced bins of 0.5~Gyr. Over plotted with a dashed curve is the Gaussian kernel density estimate of the distribution without considering the individual uncertainties in age, which neatly reproduces the main features of the underlying data distribution. When the individual fractional uncertainties in age are included in the computation of the KDE, the bimodal nature of the distribution favours a single highest peak at $\sim2$~Gyr and shows additional structure at larger ages. The optimal bandwidth in each KDE have been determined using the Sheather-Jones method that minimises the asymptotic mean integrated squared error \citep[see e.g.,][and references therein]{Sheather:1991tp,Venables2002}.

To investigate if a classical chemical selection of the disk components results in distinct age distributions, we select stars in the low- and high-$\alpha$ sequences following \citet{Adibekyan:2011ec}. We divide our sample into six metallicity bins in the range $-1.0<\mh<0.50$, and after identifying the minima in the histograms of $\afe$ distribution at each metallicity bin we draw the separation line for the components by joining the value of the minima at each bin. The result is two populations of 1030 (low-$\alpha$) and 167 (high-$\alpha$) members each, corresponding to fractions of $\sim86$\% and $\sim14$\%, respectively.
\begin{figure*}
\begin{center}
\includegraphics[width=84mm]{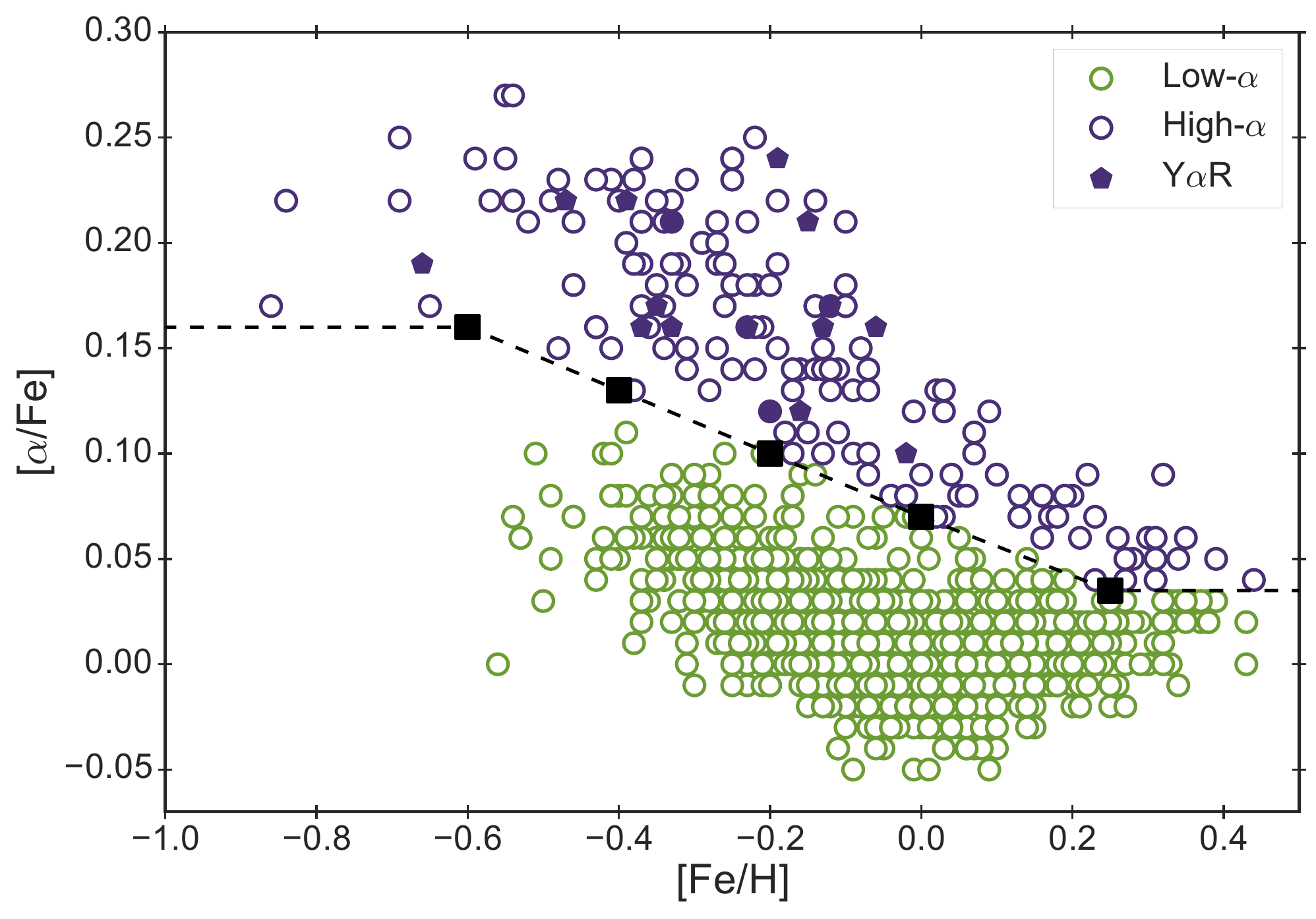}
\includegraphics[width=84mm]{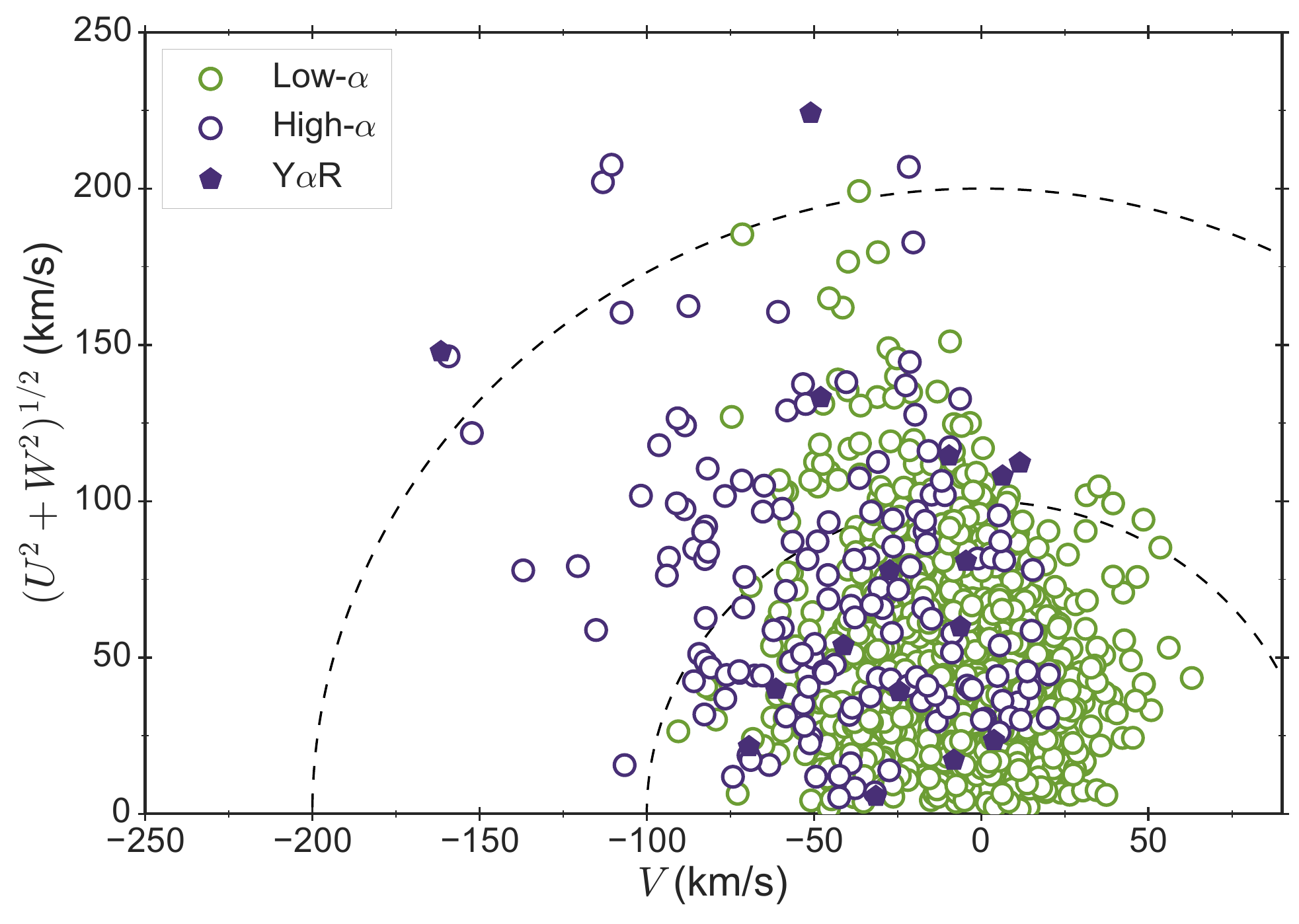}
\caption{Disk components selected based on chemistry. {\it Left}: $\alpha$-abundances as a function of metallicity. Dashed lines and black squares show the separation line between low-$\alpha$ and high-$\alpha$ disk. {\it Right}: Toomre diagram for the chemically selected components of the disk. See text for details.}
\label{fig:alp_toom_sel}
\end{center}
\end{figure*}

The kinematic and chemical properties of the sample selected in this manner are shown in Fig.~\ref{fig:alp_toom_sel}, where we also mark the position of the young $\alpha$-rich (Y$\alpha$R) stars in our sample. This population was originally identified by \citet{Chiappini:2015ja} and \citet{Martig:2015ju} as a peculiar class of stars with high $\alpha$-abundances and young ages, features not predicted by standard chemical evolution models of the Galaxy. In particular, \citet{Martig:2015ju} identified 14 stars in the APOKASC sample with $\afe>0.13$~dex and a maximum possible age younger than 6~Gyr. Only 8 of these 14 targets are contained in our colour and magnitude complete sample, and our {\tt BASTA} results confirm their apparent young nature with derived ages below 4~Gyr (consistent with the original definition of this population by  \citet{Martig:2015ju}). However, the DR13 values used in our analysis are slightly different from the DR12 spectroscopic results considered by \citet{Martig:2015ju}, and two of these 8 stars have $\alpha$-abundances slightly below the threshold of 0.13~dex (at the $\afe=0.09$ and $\afe=0.10$ level). Considering the quoted $\afe$ uncertainty of $\sim0.03$~dex in DR13, we define the young $\alpha$-rich population in our sample as those stars with $\afe>=0.10$~dex and age below 5~Gyr, obtaining a total of 16 stars fulfilling these criteria in our colour-magnitude complete sample\footnote{We have verified that a more strict cut in the selection of these young $\alpha$-rich stars does not affect our conclusions in sections \ref{ssec:ages} and \ref{ssec:dyn} regarding their origin.}.

In a similar result as found by \citet{Adibekyan:2012kr,Bensby:2014gi}, Fig.~\ref{fig:alp_toom_sel} shows that our chemically selected high-$\alpha$ and low-$\alpha$ sequences overlap drastically in kinematic space. Analogously, a separation based on their Galactic velocities leads to the opposite behaviour of overlapping in the chemical plane. We verified this by also selecting disk components based only on kinematic information following \citet{Soubiran:2003ev} and using a Gaussian Mixture Model (GMM). The rationale behind this approach is to separate the population in a finite number of Gaussian distributions based on a set of observables (which we chose to be the Galactic velocity components $U$, $V$, and $W$), and then select the number of populations according to the Bayesian Information Criterion (BIC). The model favours two Gaussian components with 1036 and 161 members respectively (fractions $\sim87$\% and $\sim13$\%), which overlap in the $\mh$ versus $\afe$ plane as expected but separate relatively well in the Toomre diagram. Nevertheless, given the probabilistic nature of the GMM some stars assigned to one Gaussian component have similar kinematics to the other Gaussian sample, making the separation between stellar populations not as clear as it is in chemical space. For this reason, we continue our analysis based on the chemical separation and explore the age dimension as an additional piece of information that can fully disentangle the components of the disk.
\subsection{The age dimension}\label{ssec:ages}
\begin{figure}
\begin{center}
\includegraphics[width=\linewidth]{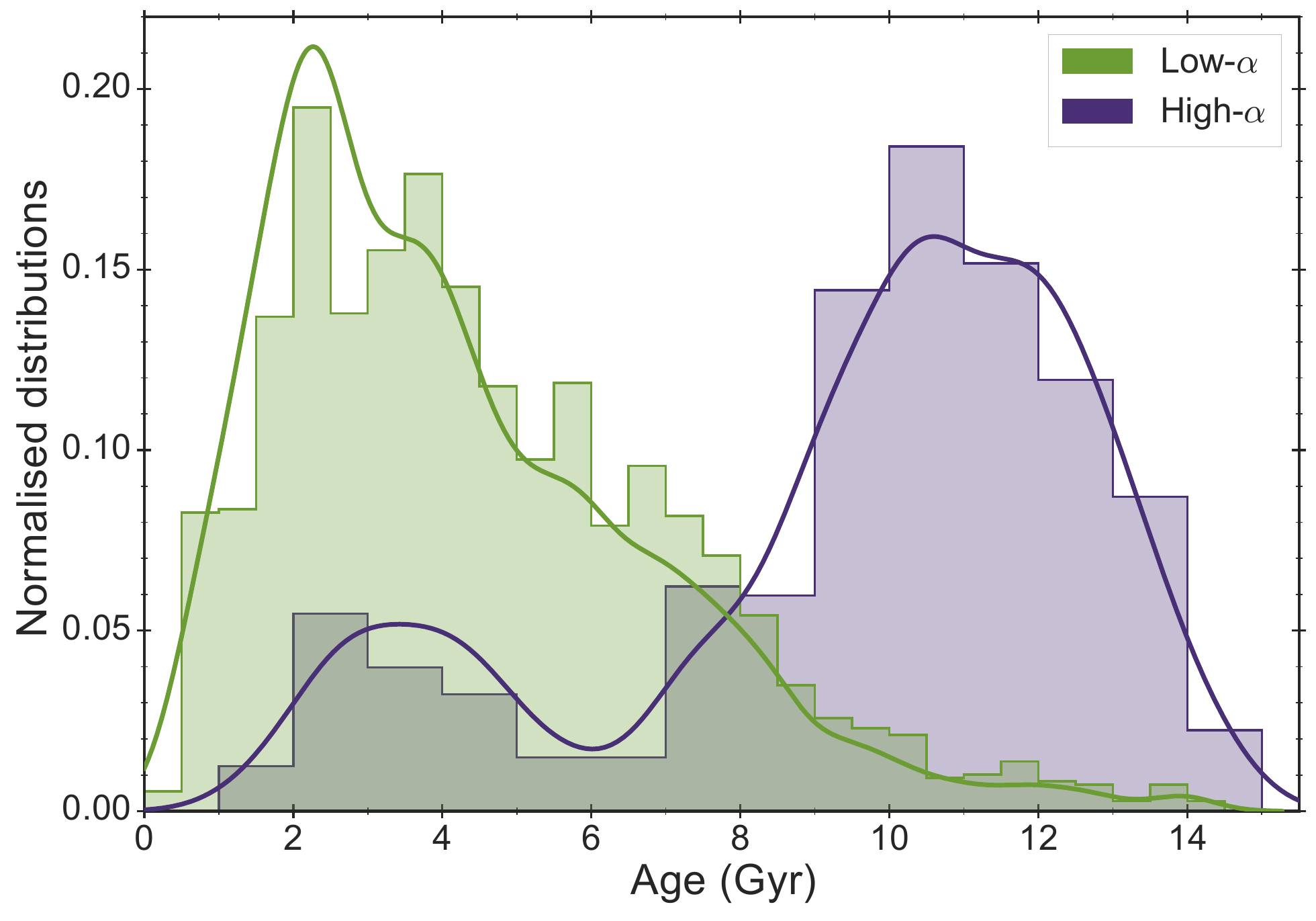}
\caption{Age distributions for the low and high-$\alpha$ disk components. The histograms are constructed using equally spaced bins of 0.5~Gyr (low-$\alpha$) and 1~Gyr (high-$\alpha$), while the solid lines represent the Gaussian KDE computed with the individual fractional uncertainties as weights and the Sheather-Jones method to determine bandwidths.}
\label{fig:age_dist}
\end{center}
\end{figure}
In Fig.~\ref{fig:age_dist} we show the age distributions of the sample selected according to the chemistry. The resulting KDE show that the low-$\alpha$ sequence peaks at $\sim2$~Gyr while the high-$\alpha$ disk does it at $\sim11$~Gyr, confirming with asteroseismology that the chemical criterion dissects the disk into two distinct populations in terms of age \citep[as shown by e.g.,][in the solar neighbourhood]{1998A&A...338..161F,Fuhrmann:2011em}. We see what appears to be contamination of old stars in the low-$\alpha$ component and young stars in the high-$\alpha$ sequence, which we further explore in the following.
\begin{figure*}
\begin{center}
\includegraphics[width=84mm]{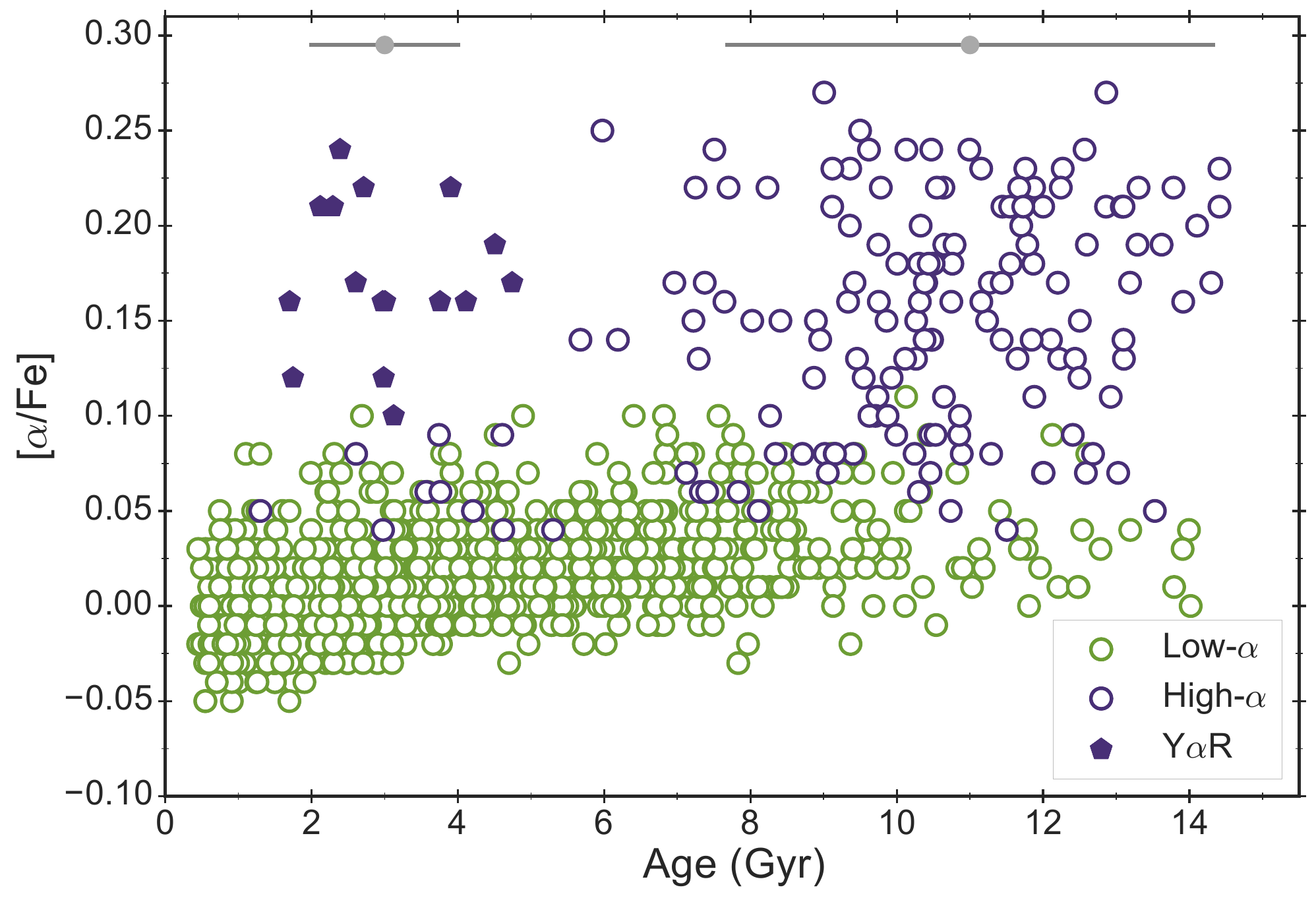}
\includegraphics[width=84mm]{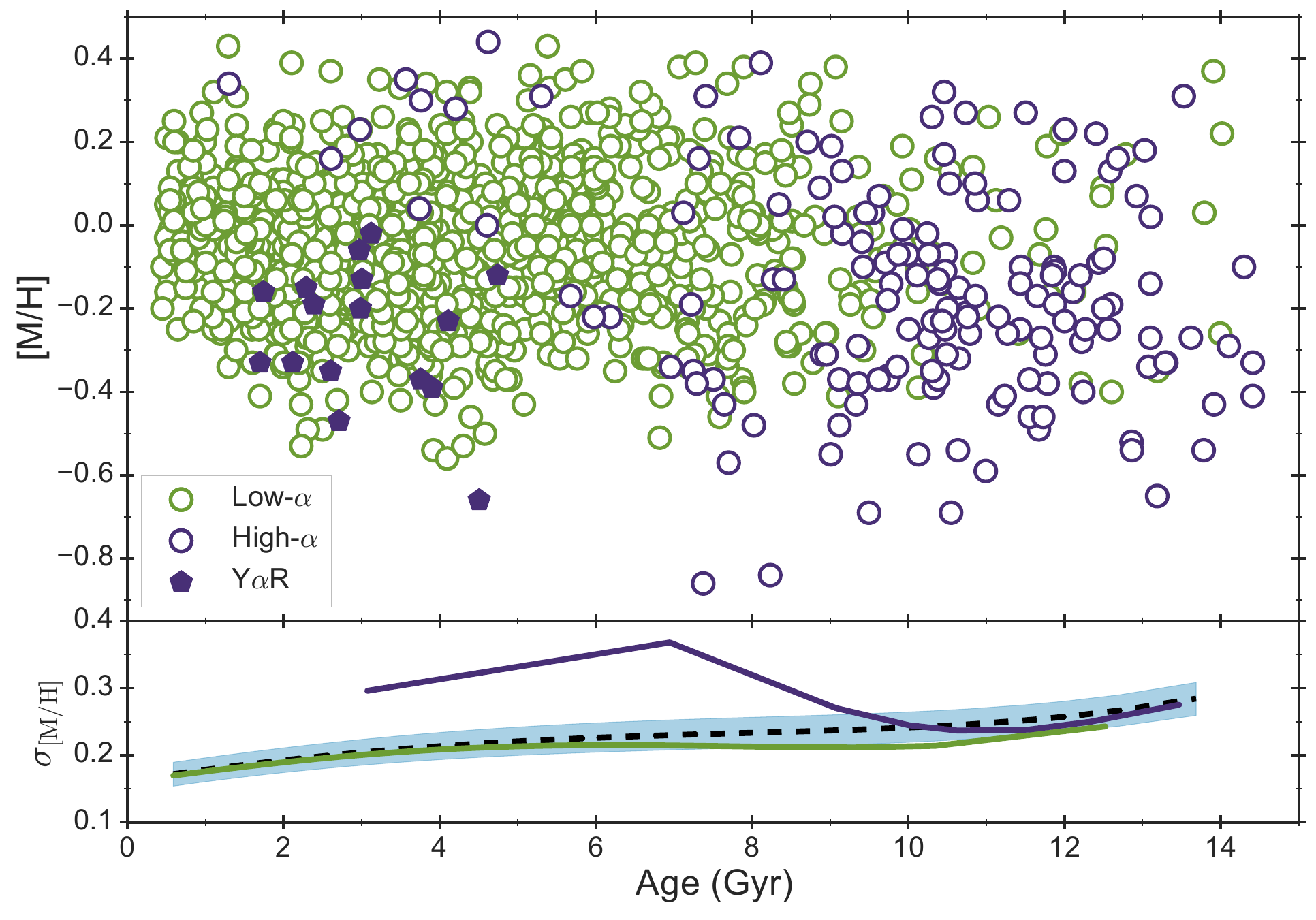}
\caption{Relations between age and chemistry for the sample of giants. {\it Left}: $\alpha$-abundances as a function of age for the chemically separated components of the disk. Grey symbols indicate median age uncertainties at 3~Gyr and 11~Gyr for the high-$\alpha$ sample. {\it Right}: Age-metallicity relation of the low-$\alpha$, and high-$\alpha$ disk populations. The bottom panel shows the metallicity dispersion determined from a bootstrap fit to the full sample (black dashed line) and its corresponding uncertainty (shaded region), as well as the high-$\alpha$ (solid violet) and low-$\alpha$ (solid green) components. See text for details.}
\label{fig:age_alph_metl}
\end{center}
\end{figure*}

The left panel in Fig.~\ref{fig:age_alph_metl} shows the relation between age and $\afe$, where it can be seen that the young $\alpha$-rich stars account for the majority of the targets in the peak at $\sim3$~Gyr seen in the high-$\alpha$ disk distribution (c.f. Fig.~\ref{fig:age_dist}). Understanding the origin of these stars has been the subject of a number of recent studies and they have been attributed to migrators from the Galactic bar \citep{Chiappini:2015ja} as well as evolved blue stragglers \citep{Martig:2015ju,Chiappini:2015ja,Yong:2016km,Jofre:2016dk}. In the former case it is believed that these stars formed in reservoirs of almost inert gas close to the end of the Galactic bar, while the latter scenario proposes that the young $\alpha$-rich stars are the product of mass transfer or stellar merger events. We further explore these scenarios using the kinematics and dynamics of Y$\alpha$R stars later in this section and in section~\ref{ssec:dyn}.

The bulk of the high-$\alpha$ stars are populating the region from $\sim8$ to $\sim14$~Gyr and show no tight correlation between age and $\afe$. In addition, old stars are not necessarily $\alpha$-rich; we identify a significant population of stars older than $\sim10$~Gyr with $\afe<0.1$ (bottom right corner in the left panel of Fig~\ref{fig:age_alph_metl}).  These findings seem in contrast to the results obtained by \citet{Haywood:2013gw}, who found a clear correlation between age and $\afe$ for both the low and high-$\alpha$ sequences and a steeper increase in $\afe$ with age for the high-$\alpha$ component. The low-$\alpha$ disk sequence on the other hand shows a gentle increase in $\afe$ with age up until $\sim8$~Gyr and remains flat after that, in agreement with the results of \citet{Haywood:2013gw}.

It is difficult to pinpoint at the moment the reason for the discrepancy between our results with those presented in \citet{Haywood:2013gw}. The authors of that paper considered a sample of 1111 solar neighbourhood dwarfs selected for exoplanet detection studies by \citet{Adibekyan:2012kr}, and determined isochrone-based ages for all stars discarding $\sim70\%$ of them due to unreliable results. Their findings are based on a subsample of only 363 stars with meaningful ages, corresponding to bright turn-off dwarfs were no assessment has been made of how representative they are of the underlaying population. Thus, the differences in our results could come from the techniques utilised (asteroseismology versus isochrone fitting), the larger number of stars in the high-$\alpha$ sequence analysed here, or the ensured completeness in our sample and appropriate correction for the target selection effects. We note, however, that other studies based on isochrones ages for turnoff stars also find larger scatter than \citet{Haywood:2013gw} in the $\alpha$ abundances of the oldest disk stars \citep[see e.g.][]{Bensby:2014gi,Bergemann:2014bm}. In the future, we expect to perform comparisons of our results with volume complete local samples to further explore the reasons for these discrepancies.

Since our sample spans distances up to 2~kpc, it is interesting to compare the predictions of chemo-dynamical models of the Milky Way with our derived properties. A diagnostic widely used in studies of the solar neighbourhood is the expected relation between age and metallicity of stars, built up as stars chemically enrich the interstellar medium. Thus, more recently formed stars should have a higher abundance of metals than those born at earlier epochs \citep[e.g.,][and references therein]{Feltzing:2001bn}. The lack of a tight relation between these parameters is attributed to e.g., the efficiency of dynamical processes capable of erasing these signatures by moving stars from their birth radii to different orbits during their lifetime \citep[see by e.g.,][]{2002MNRAS.336..785S,2008ApJ...684L..79R,Schonrich:2009ci}.

The right panel of Fig.~\ref{fig:age_alph_metl} shows the age-metallicity relation of our chemically dissected sample. Contrary to what was reported by \citet{Haywood:2013gw} we see no evidence of a significantly tighter relation between age and metallicity for the high-$\alpha$ population than for the lower-$\alpha$ sequence, but rather a broad distribution in chemical composition at all ages. Moreover, it is clear that the old (above $\sim10$~Gyr) low-$\alpha$ population shows a spread in metallicity from $-0.4<\feh<+0.4$, in disagreement with the results presented by \citet{Haywood:2013gw} where the old low-$\alpha$ stars were believed to be contaminants coming from the low-metallicity end of the low-$\alpha$ sequence with higher Galactic rotation, and hence considered to come from the outer disk. Part of this discrepancy could be explained by the age uncertainties derived for the high-$\alpha$ population in our sample (see grey symbols in Fig.~\ref{fig:age_alph_metl}), although \citet{Haywood:2013gw} quotes comparable errors of a 1.5~Gyr random component and about 1~Gyr systematic uncertainties for stars older than $\sim9$~Gyr. On the other hand sampling biases in the spectroscopic surveys can select against old low-$\alpha$ stars if these are rare \citep{Bergemann:2014bm}, and thus it is not unexpected that they appear in photometrically complete samples such as ours or the Geneva-Copenhagen survey \citep{Casagrande:2011ji}. The slope of the age-metallicity relation for the complete sample is at the level of $-0.008\pm0.001$ dex~Gyr$^{-1}$ while the dispersion\footnote{All bootstrap fits are constructed from 10,000 realisations of the dataset (with replacement) randomly drawn from a normal distribution of the mean value and uncertainty of the parameter in question, and weighted by the target selection probability of each individual target.} in the metallicity shows a clear increase as a function of age (see dashed line in the bottom-right panel of Fig.~\ref{fig:age_alph_metl}), consistent with the predictions from chemo-dynamical models including radial migration \citep[see e.g., Fig.~4 in][]{Minchev:2013ko}.

\begin{figure}
\begin{center}
\includegraphics[width=\linewidth]{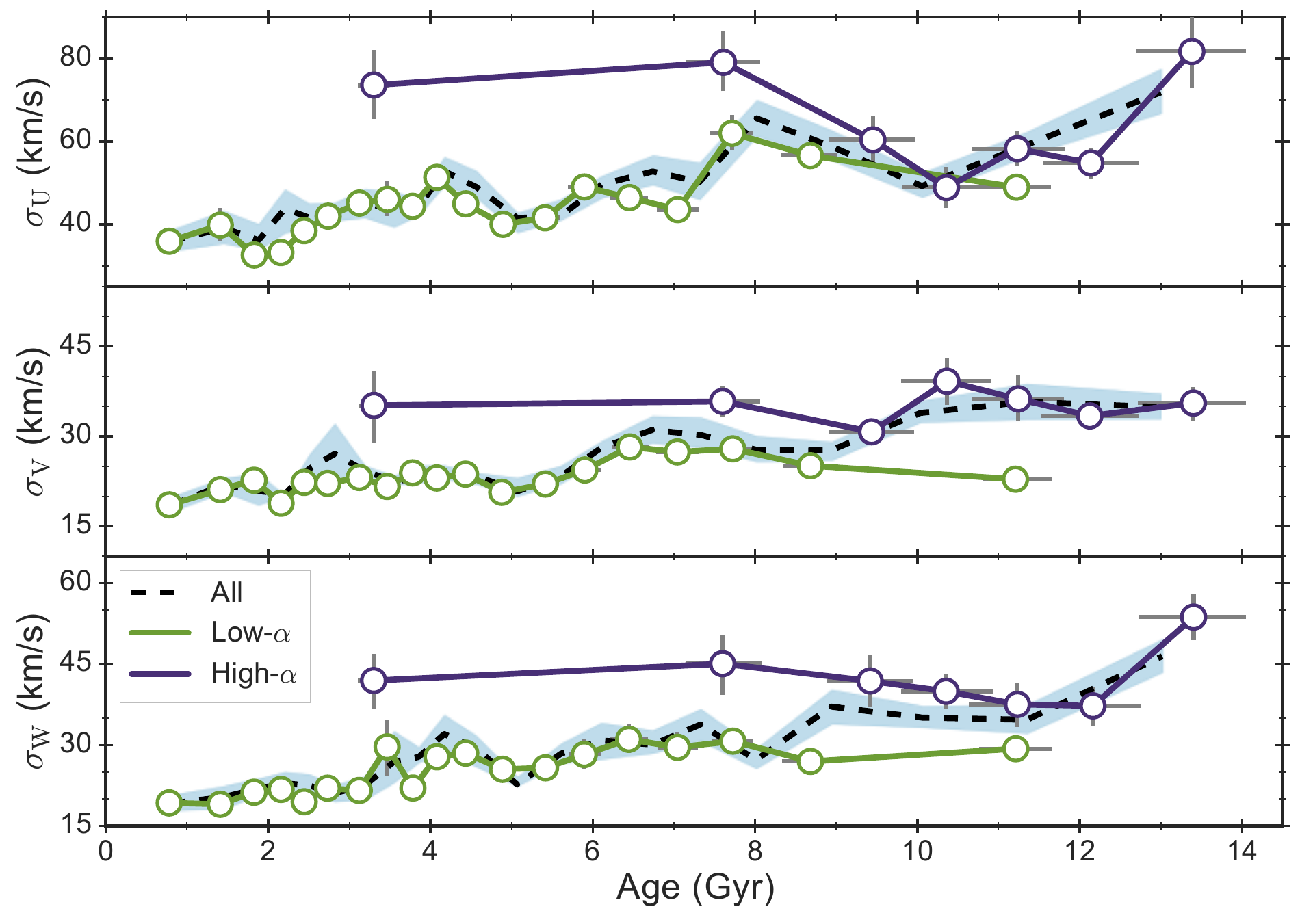}
\caption{Velocity dispersions in the $U$, $V$, and $W$ components as a function of age for all stars on the sample (dashed lines), the low-$\alpha$ (green solid line), and the high-$\alpha$ (purple solid line) sequences, determined using bootstrapping. Shaded region corresponds to bootstrap uncertainties for the full sample. See text for details.}
\label{fig:Age_veldisp}
\end{center}
\end{figure}

We now investigate variations in the stellar velocity dispersions as a function of age to disentangle kinematic signatures of different formation scenarios for the high- and low-$\alpha$ populations. Figure~\ref{fig:Age_veldisp} shows the binned distributions for these two sequences and also for the full sample. The dispersions in the three velocity components show an increase with age for the full sample, and hints of a change in slope (in $V$ and $W$) at $\sim8-9$~Gyr. This behaviour is indicative of the transition between two populations, which we explore by looking at the velocity dispersions of the individual components chemically selected in the disk.

The low-$\alpha$ sequence follows the behaviour of the full sample and steadily increases its dispersions with age until the transition point, and remains flat thereafter. The high-$\alpha$ sample presents relatively flat dispersions at all ages in the $V$ and $W$ components, and it appears to merge with the low-$\alpha$ sequence in $U$. It is interesting to see that the two chemical populations are clearly distinct in their velocity dispersions in the $V$ and $W$ components, and that the young stars belonging to the high-$\alpha$ sequence (in our definition) are kinematically hot (we remind the reader that not all of them correspond to the Y$\alpha$R population). Nevertheless, this result indicates that the young $\alpha$-rich stars have similar kinematic properties to the rest of the high-$\alpha$ sequence and are therefore likely to be born at the same time as the old high-$\alpha$ stars, rendering support to the idea that their ages are underestimated due to mass transfer or stellar merger events \citep[][]{Yong:2016km,Jofre:2016dk}. Our results indicate a $\sim10$\% fraction of young $\alpha$-rich stars among the high-$\alpha$ population, in line with the predictions by e.g., \citet{2015ApJ...807...82T} for mass transfer products \citep[see][for details]{Jofre:2016dk}.

The indication of a transition point at $\sim8-9$~Gyr in the velocity dispersions suggest that the chemically low-$\alpha$ disk formed from the initial conditions set by the high-$\alpha$ disk component. Although similar conclusions have been reached by \citet{Haywood:2013gw}, the lack of a correlation between age and $\afe$ in the high-$\alpha$ sequence argues against the scenario proposed by those authors of a quiet chemical evolution lasting 4-5~Gyr that formed the high-$\alpha$ population. In contrast, and considering the spread in bulk metallicity for stars older than $\sim8-9$~Gyr, our results indicate that over a period of more than 2~Gyr stars in the Galactic disk formed with a wide range of $\afe$ and $\mh$ independent of time. A variety of formation scenarios could produce this signature, such as chaotic gas-rich mergers occurring with inhomogeneous chemical evolution before the younger low-$\alpha$ population built up \citep[in agreement with $\Lambda$CDM predictions of decreasing merger rates, see e.g.,][]{2004ApJ...612..894B,Brook:2012ib,Minchev:2013ko,Stinson:2013dp}, or thick disk formation in large molecular clumps induced by instabilities in a gas rich disk \citep[e.g.,][]{1998Natur.392..253N,2007ApJ...670..237B,2017arXiv170807834G}. Similar results are predicted from hydrodynamical simulations where the transition point is related to for example, the decline in the star formation rate \citep[][]{2004ApJ...612..894B} and thus it corresponds to the formation period of the thick disk.
\subsection{Disk dynamics}\label{ssec:dyn}
\begin{figure}
\begin{center}
\includegraphics[width=\linewidth]{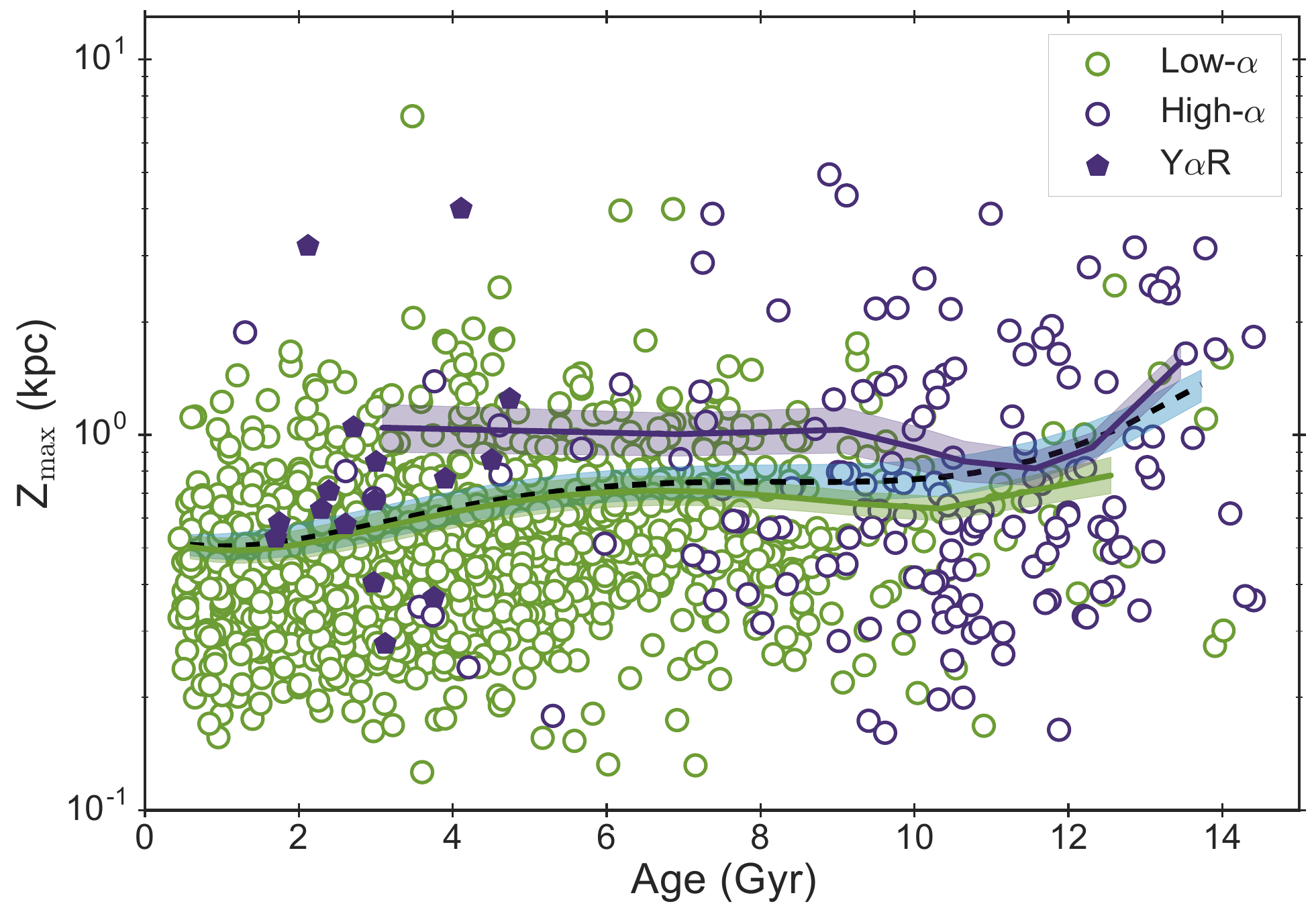}
\caption{Maximum vertical height of the stellar orbit as a function of age for the chemically dissected disk. Solid lines show the bootstrap fit and uncertainties for the mean values of each $\alpha$-selected component, while the dashed line is the same result for the full sample.}
\label{fig:Age_Zmax}
\end{center}
\end{figure}

To further explore different formation scenarios of the Milky Way disk and the origin of the two chemically defined populations, we compare the dynamical properties of our sample calculated using {\tt galpy}\footnote{http://github.com/jobovy/galpy} in the {\tt MWPotential2014} configuration \citep{Bovy:2015gg}. Figure~\ref{fig:Age_Zmax} shows the maximum vertical height of the stellar orbit as a function of age for both components, including the bootstrap fits to the mean values in both axes. The overall behaviour of the full sample is an increase in Z$_\mathrm{max}$ with age, consistent with a vertical age gradient of the disk as seen by e.g., \citet{2016ApJ...823..114N,2016MNRAS.455..987C}. The latter is normally invoked in connection with disk formations scenarios such as upside-down \citep[e.g.,][]{2006ApJ...639..126B,Bird:2013iw} and pure disk heating, or a combination thereof.

The two chemically selected sequences span a similar range of values in Z$_\mathrm{max}$ as a function of age, consistent with both populations having similar scale heights and thus not being representative of a geometric decomposition of the disk into a thin and a thick component \citep[as found by e.g.,][]{Bovy:2012ky,2016ApJ...823...30B}. The running mean for each sequence shown in Fig.~\ref{fig:Age_Zmax} reveals an average difference in Z$_\mathrm{max}$ between the populations, with some overlap at the oldest ages.

\begin{figure}
\begin{center}
\includegraphics[width=\linewidth]{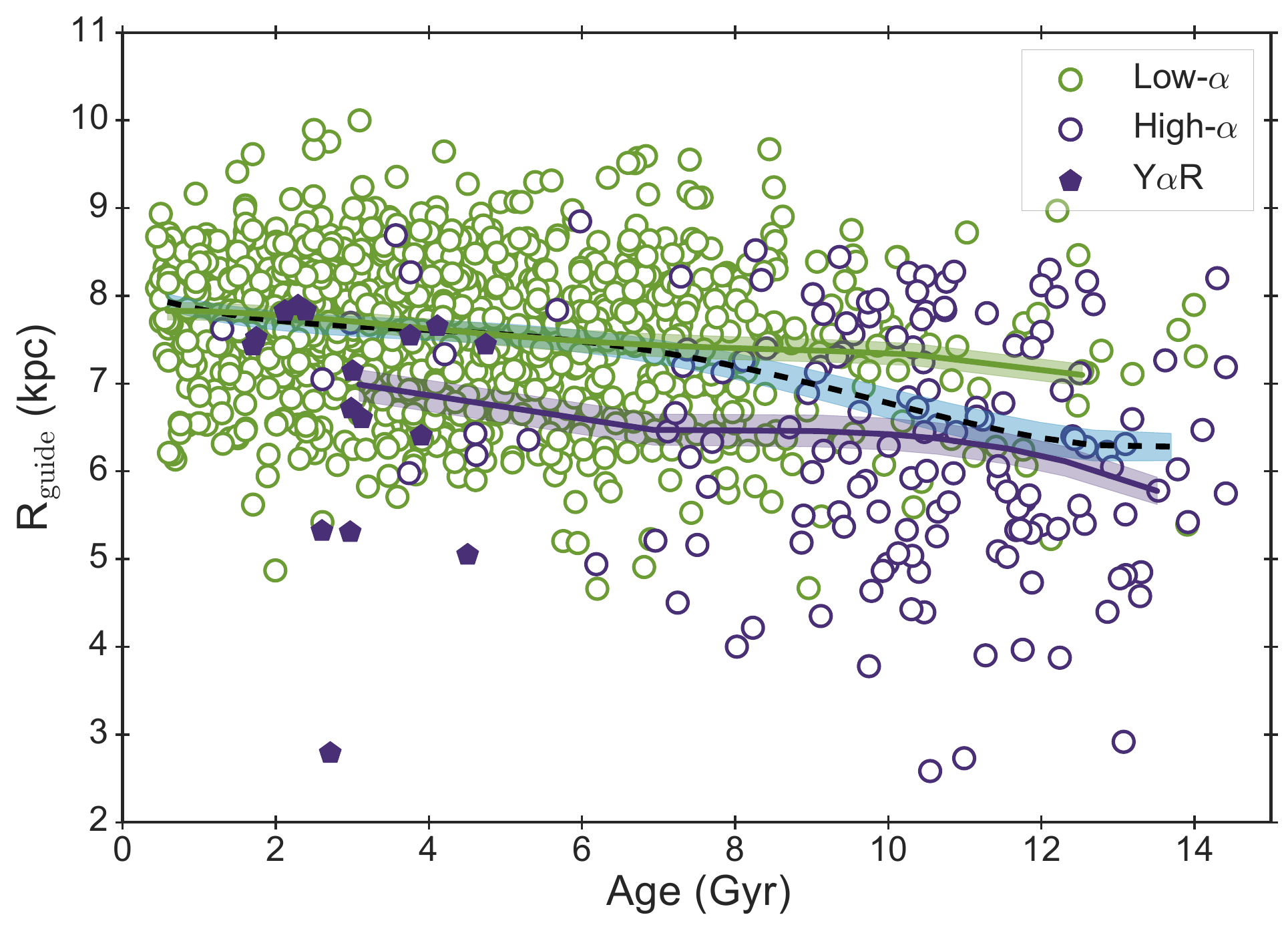}
\caption{Same as Figure~\ref{fig:Age_Zmax} for the guiding radius of the orbit.}
\label{fig:Age_Rguide}
\end{center}
\end{figure}

In Fig.~\ref{fig:Age_Rguide} we have plotted the distribution of guiding radii of the orbits, calculated as R$_\mathrm{guide}=L_z/v_c$ where $L_z$ is the angular momentum and assuming a constant circular rotation speed of $v_c=220$~km~s$^{-1}$. The running mean of R$_\mathrm{guide}$ shows a gentle decline with age for the full sample while revealing an average lower value for the high-$\alpha$ stars than for the low-$\alpha$ population \citep[see also][]{2003MNRAS.340..304R,Bensby:2011ge,Bovy:2012ky,Cheng:2012bj,AllendePrieto:2016hn}. A decrease of the overall age-R$_\mathrm{guide}$ relation can be explained by a combination of the age-velocity dispersion relation and the age-scale-length relation \citep[see e.g.,][and references therein]{TedMackereth:2017cb}. If the decrease is driven by the latter, it would indicate a smaller scale-length for the older population and thus provide direct evidence for an inside-out formation scenario of the Milky Way disk. However, an increasing age-velocity dispersion relation leads to a larger asymmetric drift that can produce a similar trend in the age-R$_\mathrm{guide}$ plane, because old stars are bound to have more negative velocities and thus come from the inner galaxy. A full chemo-dynamical model of the {\it Kepler} field is required to further explore this trend and will be the subject of a subsequent study.

Another interesting aspect from Fig.~\ref{fig:Age_Rguide} is that all young $\alpha$-rich stars have values of R$_\mathrm{guide}$ under 8~kpc, suggesting an inner Galaxy origin of this sample as postulated by \citet{Chiappini:2015ja}. However, the running mean of R$_\mathrm{guide}$ as a function of age of the high-$\alpha$ population is also below 8~Kpc, which can be indicative that the young $\alpha$-rich stars are simply a subpopulation of the high-$\alpha$ sequence with similar distribution of orbits \citep[as also proposed by][]{Martig:2015ju}. To test the null hypothesis that both the Y$\alpha$R stars and the rest of the high-$\alpha$ population are drawn from the same parent distribution, we performed 10,000 random realisations with replacement from a normal distribution constructed from the individual mean values and uncertainties in the age-R$_\mathrm{guide}$ plane for all high-$\alpha$ stars. The K-sample Anderson-Darling results show that the null hypothesis cannot be rejected with a significance level higher than 1.7\%, indicating that the guiding radius distribution of the Y$\alpha$R stars is the same as for the rest of the high-$\alpha$ component and therefore are not confined to a specific position in the inner Galaxy (e.g., close to the Galactic Bar).

\begin{figure}
\begin{center}
\includegraphics[width=\linewidth]{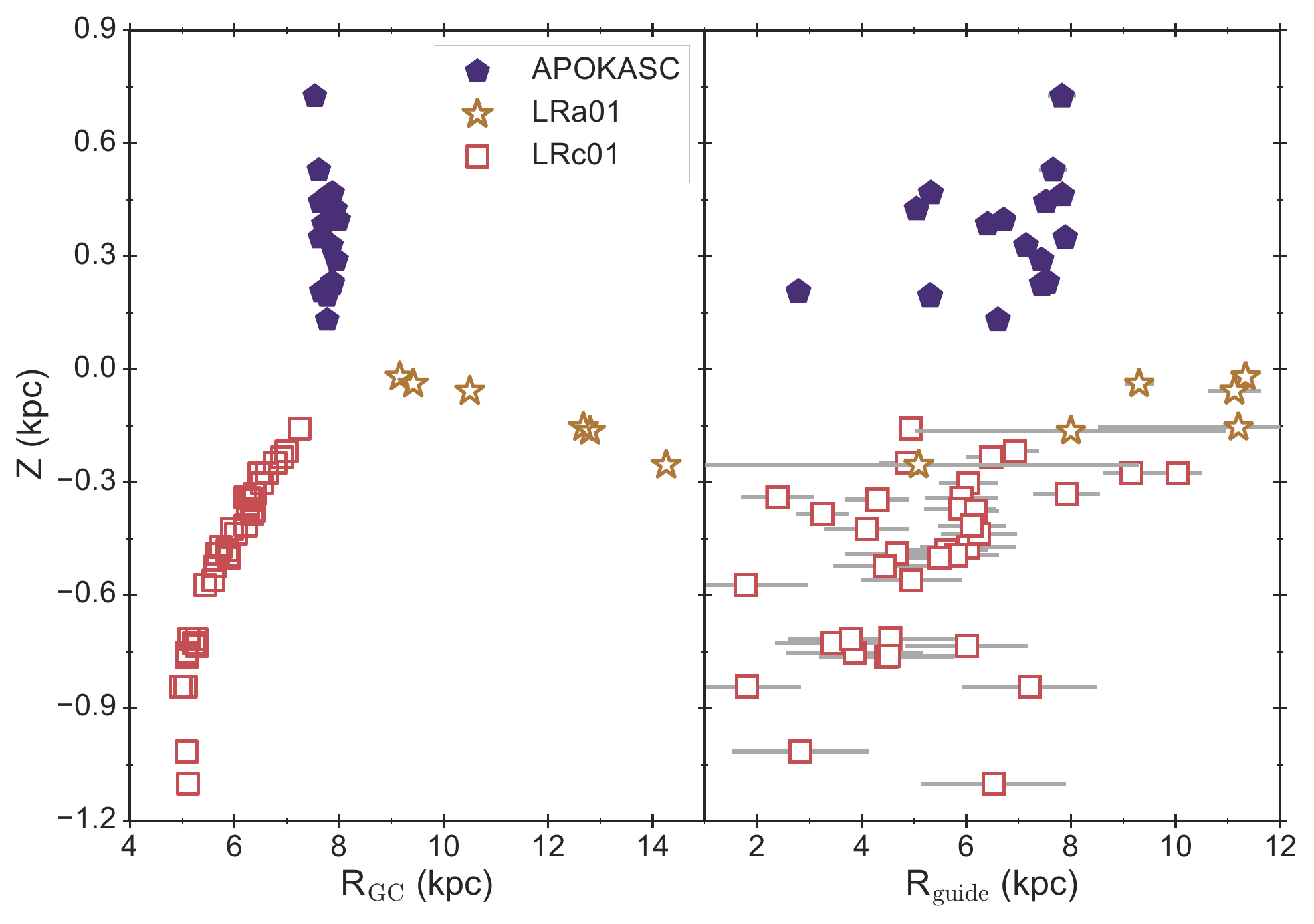}
\caption{Distribution of young $\alpha$-rich stars versus height above the plane in the Galaxy of our sample (labelled APOKASC) and the CoRoT observations (labelled LRa01 and LRc01) extracted from \citet{2017A&A...597A..30A}. {\it Left:} Galactocentric radius. {\it Right:} guiding radius and uncertainties (in most cases smaller than the symbol size for the APOKASC sample). See text for details.}
\label{fig:YAR_stars}
\end{center}
\end{figure}

The argument of an inner-Galaxy origin for the young $\alpha$-rich population was put forward by \citet{Chiappini:2015ja} after finding a larger number of these stars in the CoRoT sample looking towards the Galactic centre than in the sample observing in the anti-centre direction. One possible explanation for the discrepancy in our results could be that our {\it Kepler} stars are in the solar annulus while those of \citet{Chiappini:2015ja} are distributed across Galactocentric distances from $\sim4$ to $\sim14$~kpc. To compare the two samples we retrieved the CoRoT stellar properties published by \citet{2017A&A...597A..30A} (including ages, abundances, position in the Galaxy and guiding radii), and identified the Y$\alpha$R population using the same criteria as described in Section~\ref{sec:separ}. Figure~\ref{fig:YAR_stars} shows the spatial distribution of the CoRoT stars in the centre (LRc01) and anti-centre (LRa01) directions, accompanied by the position of our APOKASC sample.

The study by \citet{2017A&A...597A..30A} adopted proper motions from the UCAC-4 catalogue while our analysis benefited from the publication of the first Gaia data release, resulting in a much higher precision in guiding radius thanks to the improved astrometric properties (as seen in the right panel of Fig.~ \ref{fig:YAR_stars}). Despite the larger uncertainties in R$_\mathrm{guide}$ for the CoRoT stars, some general trends can be observed in the spatial distribution of the samples. Even though the CoRoT stars are also mostly confined to R$_\mathrm{guide}<8$~kpc, the higher number of Y$\alpha$R found in the inner region could be the result of probing different distances from the Galactic plane: the number of young $\alpha$-rich stars is similar towards the inner and outer part of the Galaxy for $|Z|<0.3$~kpc (see also Table~1 in \citet{Chiappini:2015ja}). Unfortunately their sample in the anti-centre direction does not reach higher values of $|Z|$ and thus can only probe larger distances from the plane towards the Galactic centre, where they find a larger number of Y$\alpha$R targets. Our sample contains 10 out of 16 young $\alpha$-rich stars at $Z>0.3$~kpc suggesting that distance to the Galactic plane instead of distance from the Galactic centre could play a role in finding an increasing number of these peculiar stars. This could simply be consequence of the increase in number of stars belonging to the high-$\alpha$ population as we move away from the plane \citep[see e.g.,][]{2015ApJ...808..132H}, thus increasing the probability of finding stars belonging to the Y$\alpha$R class. Extending the sample of these interesting stars to different Galactic directions, coupled with high precision astrometry from the Gaia mission, would help clarifying the origin of this peculiar class of objects.
\section{Conclusions}\label{sec:conc}
By combining observations from the {\it Kepler} mission with APOGEE spectroscopy, we demonstrate the power of asteroseismology as a tool for Galactic archeology and determine precise physical, chemical, and kinematic properties in a sample of more than a thousand stars fully representative of the stellar population in the direction of the {\it Kepler} field. Our main results can be summarised as follows:
\begin{itemize}
\item Thanks to asteroseismic analysis, we can confirm with high fidelity that there is a clear age difference between the low- and high-$\alpha$ components, with the low-$\alpha$ sequence peaking at $\sim2$~Gyr and the high-$\alpha$ one at $\sim11$~Gyr.
\item We observe a clear distinction in the $V$ and $W$ velocity dispersions between both components, suggesting a transition between the formation of both sequences $\sim8-9$~Gyr ago. This renders support for a formation scenario of the high-$\alpha$ component lasting more than 2~Gyr, setting the initial conditions for the evolution of the low-$\alpha$ population.
\item We see no tight correlation in $\afe$ or metallicity with age for the high-$\alpha$ sample. Our findings support a flat age metallicity relation with an increasing metallicity scatter as a function of age, consistent with models of radial migration.
\item We recover the population of seemingly young $\alpha$-rich stars found by \citet{Chiappini:2015ja,Martig:2015ju}. Their kinematics are similar to old high-$\alpha$ stars rather than low-$\alpha$ population at similar age (Fig.~\ref{fig:Age_veldisp}). This indicates that the majority of them are likely to be born at the same time as the old high-$\alpha$ stars, and therefore supports the idea that they are stellar merger remnants \citep{Yong:2016km,Jofre:2016dk}. We also find evidence that the young $\alpha$-rich stars are vertically hotter than the low-$\alpha$ population (Fig~\ref{fig:Age_Zmax}), and that their guiding radii follows the same distribution as the rest of the high-$\alpha$ sequence. This seems to be inconsistent with previous interpretations of the spacial distribution of these young $\alpha$-rich stars  as coming from the inner part of the Galaxy, suggesting that they formed in the bar region and migrated outwards. The latter scenario could be the result of an incomplete sampling at distances larger than $|Z|>0.3$~kpc in the anti-centre direction in previous studies (see Fig~\ref{fig:YAR_stars}).
\item The distribution of Z$_\mathrm{max}$ and R$_\mathrm{guide}$ as a function of age for the full stellar sample analysed could provide evidence of inside-out formation of the Milky Way disk. A complete chemo-dynamical simulation of our sample in the {\it Kepler} field could help disentangle between this scenario and the effects of the asymmetric drift.
\end{itemize}

It is quite interesting that we find many old high metallicity and low-$\alpha$ stars, which are not seen in the solar neighbourhood studies \citep{Haywood:2013gw,Bensby:2014gi} but are predicted by standard chemical evolution models \citep[see e.g., Fig~15. in][]{Nidever:2014fj}. This may be due to the different methods for age determination (asteroseismology versus isochrone fitting) or could also be related to the difference in sample selection: we have ensured a representative group of stars to distances up to 3~kpc from us while the solar neighbourhood studies are mostly defined by much stricter data availability. A detailed comparison between ages determined from isochrone fitting and asteroseismic inferences could help shed some light on this topic. Similarly, data at different galactocentric radii and height from the plane would serve for studying how the age-$\afe$ and age-metallicity relations change at different location of the disk. Future analysis based on K2 for Galactic archaeology \citep[][]{2015ApJ...809L...3S} can probe many other directions and thus extend our findings to other regions of the Milky Way, while a complete revision of the solar neighbourhood sample based on asteroseismic data will be possible from the TESS mission \citep{Ricker:2015ie}.
\section*{Acknowledgements}
We would like to thank the anonymous referee for carefully revising the manuscript and suggesting changes that have improved the quality of the paper. Funding for this Discovery mission is provided by NASA's Science Mission Directorate. The authors acknowledge the dedicated team behind the Kepler and K2 missions, without whom this work would not have been possible. Funding for the Stellar Astrophysics Centre is provided by The Danish National Research Foundation (Grant agreement No.~DNRF106). V.S.A. acknowledges support from VILLUM FONDEN (research grant 10118). L.C. is the recipient of an Australian Research Council Future Fellowship (project number FT160100402). D.K. and I.C. acknowledges the support of the UK's Science \& Technology Facilities Council (STFC Grant ST/N000811/1 and Doctoral Training Partnerships Grant ST/N504488/1). MNL acknowledges the support of The Danish Council for Independent Research | Natural Science (Grant DFF-4181-00415). D.H. acknowledges support by the Australian Research Council's Discovery Projects funding scheme (project number DE140101364) and support by the National Aeronautics and Space Administration under Grant NNX14AB92G issued through the Kepler Participating Scientist Program. J.A.J., M.P., and J.T. acknowledge support from NSF Grant AST-1211673. A.S. acknowledges support from grant ESP2015-66134-R (MINECO). D.S. is the recipient of an Australian Research Council Future Fellowship (project number FT1400147). W.H.T. acknowledges funding from the European Research Council under the European Union's Seventh Framework Programme (FP 7) ERC Grant Agreement n.~${\rm [321035]}$.

Funding for the Sloan Digital Sky Survey IV has been provided by the Alfred P. Sloan Foundation, the U.S. Department of Energy Office of Science, and the Participating Institutions. SDSS acknowledges support and resources from the Center for High-Performance Computing at the University of Utah. The SDSS web site is www.sdss.org. SDSS is managed by the Astrophysical Research Consortium for the Participating Institutions of the SDSS Collaboration including the Brazilian Participation Group, the Carnegie Institution for Science, Carnegie Mellon University, the Chilean Participation Group, the French Participation Group, Harvard-Smithsonian Center for Astrophysics, Instituto de Astrofísica de Canarias, The Johns Hopkins University, Kavli Institute for the Physics and Mathematics of the Universe (IPMU) / University of Tokyo, Lawrence Berkeley National Laboratory, Leibniz Institut f\"ur Astrophysik Potsdam (AIP), Max-Planck-Institut f\"ur Astronomie (MPIA Heidelberg), Max-Planck-Institut f\"ur Astrophysik (MPA Garching), Max-Planck-Institut f\"ur Extraterrestrische Physik (MPE), National Astronomical Observatories of China, New Mexico State University, New York University, University of Notre Dame, Observatório Nacional / MCTI, The Ohio State University, Pennsylvania State University, Shanghai Astronomical Observatory, United Kingdom Participation Group, Universidad Nacional Autónoma de México, University of Arizona, University of Colorado Boulder, University of Oxford, University of Portsmouth, University of Utah, University of Virginia, University of Washington, University of Wisconsin, Vanderbilt University, and Yale University.
\footnotesize{
\bibliographystyle{mn2e}
\bibliography{diskage}
%}

%
\appendix
\section{Stellar properties derived with {\tt BASTA}}\label{app_props}
The stellar properties derived with {\tt BASTA} for the sample of stars analysed in this paper are published in the online version of the article. A description of all fields available is presented in Table~\ref{tab:fin_props}.
\begin{table}
\centering
\caption{Stellar properties determined with {\tt BASTA} for the APOKASC sample.}
\label{tab:fin_props}
\begin{tabular}{ll}
\hline
\hline
Field & Description \\
\hline
{\ttfamily KIC} & {\it Kepler} Input Catalogue Identifier\\
{\ttfamily Mass} & Mass in solar units \\
{\ttfamily Mass\_err} & Mass uncertainty in solar units \\
{\ttfamily Rad} & Radius in solar units \\
{\ttfamily Rad\_err} & Radius uncertainty in solar units \\
{\ttfamily Grav} & Surface gravity in dex \\
{\ttfamily Grav\_err} & Surface gravity uncertainty in dex \\
{\ttfamily Age} & Age in units of Gyr \\
{\ttfamily Age\_err} & Age uncertainty in units of Gyr \\
{\ttfamily Lum} & Luminosity in solar units \\
{\ttfamily Lum\_err} & Luminosity uncertainty in solar units \\
{\ttfamily Dist} & Distance in pc \\
{\ttfamily Dist\_err} & Distance uncertainty in pc \\
{\ttfamily Prob} & Target selection probability (cf section~\ref{sec:tarprob}) \\
\hline
\end{tabular}
\end{table}
\section{Tests of target selection}\label{app_tsel}
Our procedure for correcting for target selection effects outlined in section~\ref{sec:tarprob} assumes flat priors in age, metallicity and distances. In fact, the elegance of the data-cube approach is to return the probability of observing a star at any given point of the parameter space without invoking any Galactic modelling. The only assumption entering our method is the IMF, which we adopt to be a Salpeter one. We have tested the effect of drastically changing this prior, by assuming a uniform distribution in mass (flat IMF), and generating a new multi-dimensional data cube to determine the probability of detection. Figure~ \ref{fig:targ_selFlatIMF} shows the resulting normalised probability as a function of age, metallicity, and distance, which can be directly compared to Fig.~\ref{fig:targ_sel}. The overall shape of the distributions is unchanged.
\begin{figure}
\begin{center}
\includegraphics[width=\linewidth]{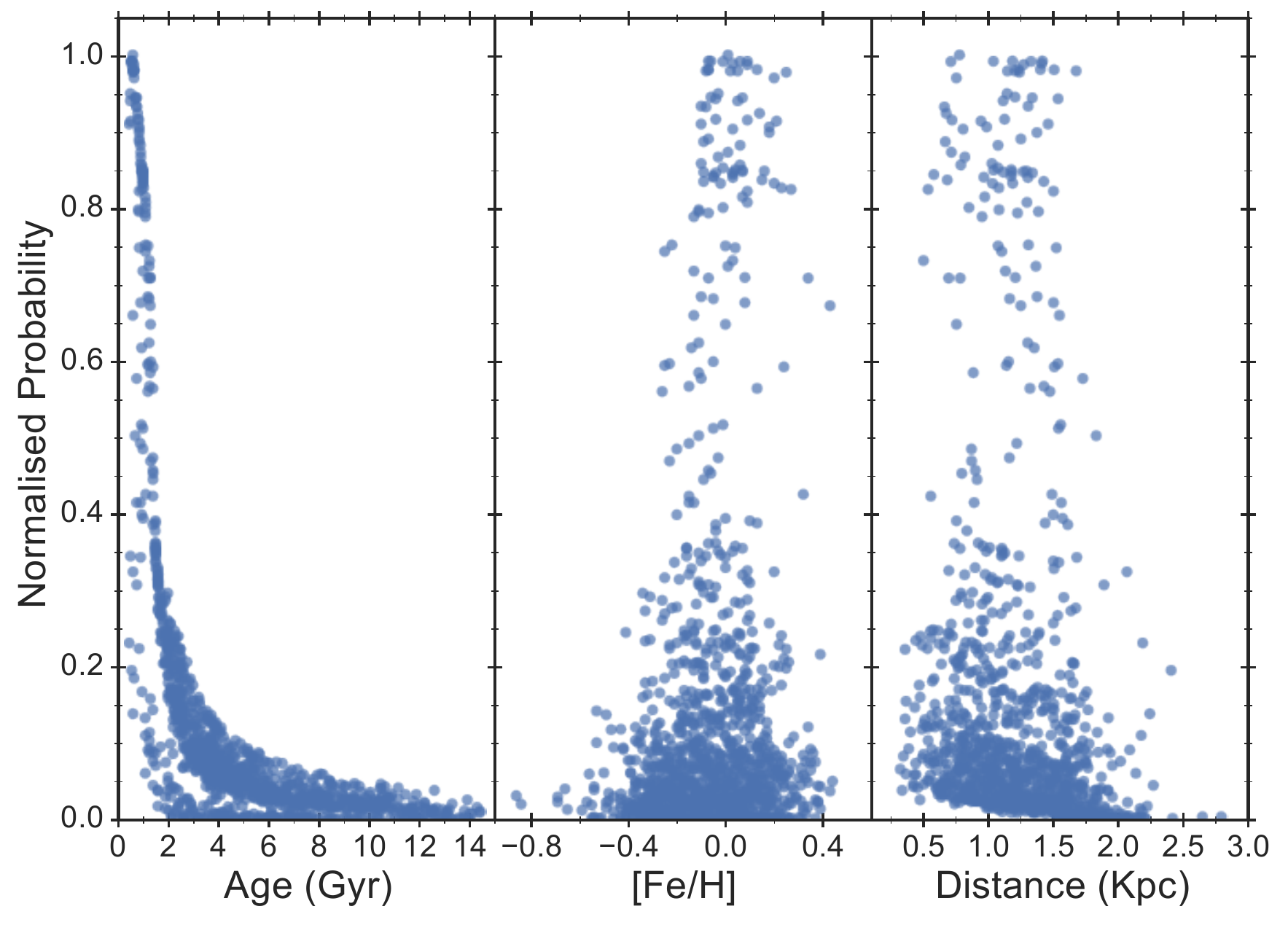}
\caption{Same as Fig~\ref{fig:targ_sel} but applying a correction for target selection based on a uniform prior in the IMF.}
\label{fig:targ_selFlatIMF}
\end{center}
\end{figure}

Figure~\ref{fig:age_distFlatIMF} shows the age distribution of the low- and high-$\alpha$ samples corrected for target selection using the probabilities calculated with a uniform mass prior. As it can be see the resulting distributions are almost unchanged compared to those shown in Fig.~\ref{fig:age_dist}, also peaking at values of $\sim2$ and $\sim11$~Gyr for the low- and high-$\alpha$ components.
\begin{figure}
\begin{center}
\includegraphics[width=\linewidth]{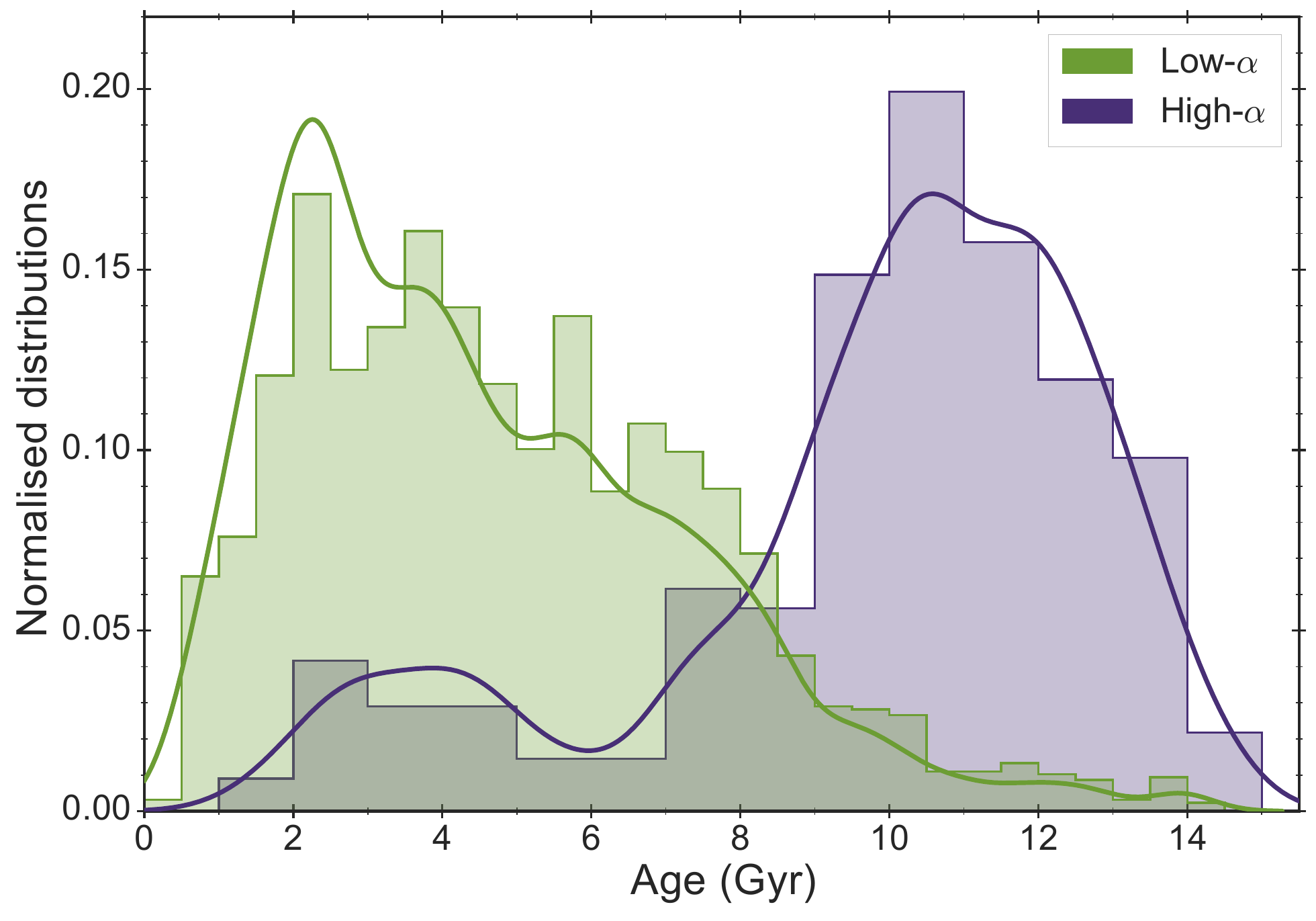}
\caption{Same as Fig~\ref{fig:age_dist} but applying a correction for target selection based on a uniform prior in the IMF.}
\label{fig:age_distFlatIMF}
\end{center}
\end{figure}
\end{document}